\documentclass[prd,twocolumn,superscriptaddress,showpacs,nofootinbib,preprintnumbers]{revtex4}

\usepackage{amsmath}
\usepackage{amsfonts}
\usepackage{graphicx}
\usepackage{dcolumn}
\def\hs{\qquad}               
\def\beq{\begin{eqnarray}}    
\def\eeq{\end{eqnarray}}      
\def\ap{\left.}               
\def\at{\left(}               
\def\aq{\left[}               
\def\cp{\right.}              
\def\ct{\right)}              
\def\cq{\right]}              
\def\Tr{\mathop{\rm Tr}\nolimits}                  
\def\lap{\Delta\,}                                 
\def\al{\alpha}

\def\ka{\kappa}

\def\si{\sigma}

\def\Ga{\Gamma}
\def\De{\Delta}

\newcommand{\be}{\begin{equation}}
\newcommand{\ee}{\end{equation}}
\newcommand{\bea}{\begin{eqnarray}}
\newcommand{\eea}{\end{eqnarray}}
\newcommand{\beaa}{\begin{eqnarray*}}
\newcommand{\eeaa}{\end{eqnarray*}}

\newcommand{\nn}{\nonumber \\}
\newcommand{\e}{{\rm e}}
\begin{document}
\tolerance=5000

\title{Dark energy in modified Gauss-Bonnet gravity:
late-time acceleration and the hierarchy problem}
\author{Guido Cognola$\,^{(a)}$\footnote{cognola@science.unitn.it},
Emilio Elizalde$\,^{(b)}$\footnote{elizalde@ieec.uab.es},
Shin'ichi Nojiri$\,^{(c)}$\footnote{nojiri@nda.ac.jp,
snojiri@yukawa.kyoto-u.ac.jp},\\
Sergei D.~Odintsov$\,^{(b,d)}$\footnote{odintsov@ieec.uab.es also at TSPU,
Tomsk}
and
Sergio Zerbini$\,^{(a)}$\footnote{zerbini@science.unitn.it}
}
\affiliation{
$^{(a)}$ Dipartimento di Fisica, Universit\`a di Trento \\
and Istituto Nazionale di Fisica Nucleare \\
Gruppo Collegato di Trento, Italia\\
\medskip
$^{(b)}$ Consejo Superior de Investigaciones Cient\'{\i}ficas
(ICE/CSIC) \, and \\  Institut d'Estudis Espacials de Catalunya
(IEEC) \\
Campus UAB, Fac Ciencies, Torre C5-Par-2a pl \\ E-08193 Bellaterra
(Barcelona) Spain\\
\medskip
$^{(c)}$ Department of Applied Physics,
National Defence Academy, \\
Hashirimizu Yokosuka 239-8686, Japan \\
\medskip
$^{(d)}$ ICREA, Barcelona, Spain \, and \\
Institut d'Estudis Espacials de Catalunya (IEEC) \\
Campus UAB, Fac Ciencies, Torre C5-Par-2a pl \\ E-08193 Bellaterra
(Barcelona) Spain\\
}

\begin{abstract}

Dark energy cosmology is considered in a modified Gauss-Bonnet (GB) model
of gravity where an arbitrary function of the GB invariant, $f(G)$,  is
added to the General Relativity action. We show that a theory of this kind
is endowed with a quite rich cosmological structure: it may naturally 
lead to an effective cosmological constant, quintessence or phantom cosmic
acceleration, with a possibility for the transition from deceleration 
to acceleration.
It is demonstrated in the paper that this theory is perfectly viable,
since it is compliant with the Solar System constraints. 
Specific properties
of $f(G)$ gravity in a de Sitter universe, such as dS and SdS solutions,
their entropy and its explicit one-loop quantization are studied. The
issue of a possible solution of the hierarchy problem in modified
gravities is addressed too.

\end{abstract}

\pacs{98.70.Vc}

\maketitle

\section{Introduction}

Recent observational data indicate that our universe is
accelerating. This acceleration is explained in terms of the
so-called dark energy (DE) which could result from a cosmological
constant, an ideal fluid with a (complicated) equation of state and
negative pressure, the manifestation of vacuum effects, a scalar (or
more sophisticated) field, with quintessence-like or phantom-like
behavior, etc. (For a very complete review of a dynamical DE see
\cite{sami} and references therein, for an earlier review, see
\cite{pad}.) The choice of possibilities
reflects the undisputable fact that the true nature and origin of 
the dark energy
has not been convincingly explained yet. It is not even clear what
type of DE (cosmological constant, quintessence or phantom) occurs
in the present, late universe.

A quite appealing possibility for the gravitational origin of the DE is
the modification of General Relativity. Actually, there is no
compelling reason why standard GR should be trusted at large cosmological
scales. For a rather minimal modification, one assumes that the
gravitational action may contain some additional terms which start
to grow slowly with decreasing curvature (of type $1/R$
\cite{capozziello,NOPRD}, $\ln R$ \cite{ln}, $\Tr 1/R$ \cite{tr},
string-inspired dilaton gravities \cite{string}, etc.), and which
could be responsible for the current accelerated expansion. In
fact, there are stringent constraints on these apparently harmless
modifications of General Relativity coming from precise Solar System
tests, and thus not many of these modified gravities may be viable
in the end. In such situation, a quite natural explanation for both
the cosmic speed-up issue and also of the first and second
coincidence problems (for recent discussion of the same, see
\cite{cai}) could be to say that all of them are caused, in fact, by the
universe expansion itself! Nevertheless, one should not forget that
some duality exists between the ideal fluid equation of state (EoS)
description, the scalar-tensor theories and modified gravity
\cite{CNO}. Such duality leads to {\it the same} FRW dynamics,
starting from three physically different ---but mathematically
equivalent--- theories. Moreover, even for modified gravity,
different actions may lead to the same FRW dynamics\cite{multamaki}.
Hence, additional evidence in favor of one or another DE model (with
the same FRW scale factor) should be clearly exhibited \cite{CNO}.

As a simple example, let us now see how different types of DE may
actually show up in different ways at large distances. It is well
known that cold dark matter is localized near galaxy clusters but,
quite on the contrary, dark energy distributes uniformly in the
universe. The reason for that could be explained by a difference in
the EoS parameter $w=p/\rho$. As we will see in the following, the
effect of gravity on the cosmological fluid depends on $w$ and even
when $-1<w<0$ gravity can act sometimes as a repulsive force.

To see the $w-$dependence on the fluid distribution in a quite
simple example, we consider cosmology in anti-de Sitter (AdS) space,
whose metric is given by \be \label{AdS1} ds^2 = dy^2 +
\e^{2y/l}\left(-dt^2 + \sum_{i=1,2}\left(dx^i\right)^2\right)\ , \ee
the conservation law of the energy momentum tensor, $\nabla_\mu
T^{\mu\nu}$ gives, by putting $\nu=y$, \be \label{AdS2}
\frac{dp}{dy} + \frac{1}{l}\left(p+\rho\right)=0\ , \ee if we assume
matter to depend only on the coordinate $y$. When $w$ is a constant,
we can solve Eq.~(\ref{AdS2}) explicitly \be \label{AdS3}
\rho=\rho_0\exp\left[ -
\frac{1}{l}\left(1+\frac{1}{w}\right)y\right]\ . \ee Here $\rho_0$
is a constant. We should note that $1 + 1/w>0$ when $w>0$ or $w<-1$,
and $1 + 1/w <0$ when $-1<w<0$. Then, for usual matter with $w>0$,
the density $\rho$ becomes large when $y$ is negative and large. In
particular, for dust $w\sim 0$ but $w>0$, and a collapse would
occur. On the other hand, when $-1<w<0$, like for quintessence, the
density $\rho$ becomes large when $y$ is positive. In the phantom
case, with $w<-1$, the density $\rho$ becomes large when $y$ is
negative, although a collapse does not occur. When $w=-1$, $\rho$
becomes constant and uniform.

We may also consider a Schwarzschild like metric: \be \label{Sc1}
ds^2 = - \e^{2\nu(r)}dt^2 + \e^{-2\nu(r)}dr^2 + r^2 d\Omega_{(2)}^2\
. \ee Here $d\Omega_{(2)}^2$ expresses the metric of a
two-dimensional sphere of unit radius. Then, the conservation law
$\nabla_\mu T^{\mu\nu}$ with $\nu=r$ gives, \be \label{Sc2}
\frac{dp}{dr} + \frac{d\nu}{dr}\left(p+\rho\right)=0\ . \ee When $w$
is constant, we can solve (\ref{Sc2}) and obtain \be \label{Sc3}
\rho=\rho_0\exp\left[ - \left(1+\frac{1}{w}\right)\nu\right]\ . \ee
Here $\rho_0$ is a constant again. In particular, in the case of the
Schwarzschild metric, \be \label{Sc4} \e^{2\nu(r)}=1 -
\frac{r_0}{r}\ , \ee with horizon radius $r_0$, we find \be
\label{Sc5} \rho=\rho_0\left(1-\frac{r_0}{r}\right)^{ -
\frac{1}{2}\left(1+\frac{1}{w}\right)}\ . \ee Then, when $w>0$ or
$w<-1$, $\rho$ is a decreasing function of $r$, that is, the fluid
is localized near the horizon. Specifically, in the case of dust
with $w=0$, the fluid collapses. On the other hand, when $-1<w<0$,
$\rho$ is an increasing function of $r$, which means that the fluid
delocalizes. When $w=0$, the distribution of the fluid is uniform.

The above results tell us that the effect of gravity on matter
with $-1<w<0$ is opposite to that on usual matter. Usual matter
becomes dense near a star but matter with $-1<w<0$ becomes less
dense when approaching a star. As is known, cold dark matter
localizes near galaxy clusters but dark energy distributes uniformly
within the universe, which would be indeed consistent, since the EoS
parameter of dark energy is almost $-1$. If dark energy is of
phantom nature ($w<-1$), its density becomes large near the cluster
but if dark energy is of quintessence type ($-1<w<-1/3$), its
density becomes smaller.

In the present paper the (mainly late-time) cosmology coming from
modified Gauss-Bonnet (GB) gravity, introduced in Ref.\cite{GB}, is
investigated in detail. In the next section, general FRW equations
of motion in modified GB gravity with matter are derived. Late-time
solutions thereof, for various choices of the function $f(G)$, are
found. It is shown that modified GB gravity may indeed play the role
of a gravitational alternative for DE. In particular, we demonstrate
that this model may naturally lead to a plausible, effective
cosmological constant, quintessence or a phantom era. In addition,
 $f(G)$ gravity has the possibility to describe the inflationary era 
(unifying then inflation with late-time acceleration), and to yield 
a transition from
deceleration to acceleration, as well as a natural crossing of the
phantom divide. It also passes the stringent solar system tests, as
it shows no correction to Newton's law in flat space for an
arbitrary choice of $f(G)$, as well as no instabilities. Sect. 3 is
devoted to the study of the de Sitter universe solution in such
model. The entropies of a SdS black hole and of a dS universe are
derived, and possible applications to the calculation of the
nucleation rate are discussed. In Sect. 4, the quantization program
at one-loop order for modified GB gravity is presented. This issue
is of the essence for the phantom era, where quantum gravity effects
eventually become important near the Big Rip singularity. Sect. 5 is
devoted to the generalization of modified gravity where $F=F(G,R)$.
This family of models looks less attractive, given that only some of its
specific realizations may pass the solar system tests. Nevertheless, it
can serve to discuss the origin of the cosmic speed-up as well as a
possible transition from deceleration to acceleration. In Sect. 6,
the important hierarchy problem of particle physics is addressed in
the framework of those modified gravity theories. It is demonstrated
there that this issue may have a natural solution in the frame of
${\cal F}(R)$ or $F(G,R)$-gravity. The last section
is devoted to a summary and an outlook. In the Appendix, an attempt is 
made to construct zero curvature black hole solutions in the theory under
discussion.

\section{Late-time cosmology in modified Gauss-Bonnet gravity}

Let us start from the following, quite general action for modified gravity 
\cite{BT}
\be
\label{GB0}
S=\int d^4x\sqrt{-g} \left(\tilde f\left( R, R_{\mu\nu}R^{\mu\nu},
R_{\mu\nu\rho\sigma}R^{\mu\nu\rho\sigma} \right) + {\cal L}_m\right)\ .
\ee
Here ${\cal L}_m$ is the matter Lagrangian density.
It is not easy to construct a viable theory directly from this general
class, which allows for non-linear forms for the action.
One must soon make use of symmetry considerations, which lead to theories 
which are more friendly, e.g., to the common Solar System tests.
Specifically, we shall restrict the action to the following form:
\be
\label{GR1}
S=\int d^4 x\sqrt{-g}\left(F(G,R) + {\cal L}_m\right)\ .
\ee
Here $G$ is the GB invariant:
\begin{equation}
G=R^2 -4 R_{\mu\nu} R^{\mu\nu} + R_{\mu\nu\xi\sigma}R^{\mu\nu\xi\sigma}\,.
\label{GB}
\end{equation}
Varying over $g_{\mu\nu}$:
\bea
\label{GR2}
&& 0= T^{\mu\nu} + \frac{1}{2}g^{\mu\nu} F(G,R) \nn && -2 F_G(G,R) R R^{\mu\nu}
+
4F_G(G,R)R^\mu_{\ \rho} R^{\nu\rho} \nn
&& -2 F_G(G,R) R^{\mu\rho\sigma\tau}R^\nu_{\ \rho\sigma\tau}
    -4 F_G(G,R) R^{\mu\rho\sigma\nu}R_{\rho\sigma} \nn
&& + 2 \left( \nabla^\mu \nabla^\nu F_G(G,R)\right)R
    - 2 g^{\mu\nu} \left( \nabla^2 F_G(G,R)\right)R \nn
&& - 4 \left( \nabla_\rho \nabla^\mu F_G(G,R)\right)R^{\nu\rho}
    - 4 \left( \nabla_\rho \nabla^\nu F_G(G,R)\right)R^{\mu\rho} \nn
&& + 4 \left( \nabla^2 F_G(G,R) \right)R^{\mu\nu} + 4g^{\mu\nu}
\left( \nabla_{\rho} \nabla_\sigma F_G(G,R) \right) R^{\rho\sigma} \nn
&& - 4 \left(\nabla_\rho \nabla_\sigma F_G(G,R) \right)
R^{\mu\rho\nu\sigma} \nn && - F_R(G,R) R^{\mu\nu} + \nabla^\mu
\nabla^\nu F_R(G,R) \nn && - g^{\mu\nu}\nabla^2 F_R(G,R) \ ,
\eea
$T^{\mu\nu}$ being the matter-energy momentum tensor, and where
the following expressions are used:
\be
\label{GR3}
F_G(G,R)=\frac{\partial F(G,R)}{\partial G}\ ,\quad
F_R(G,R)=\frac{\partial F(G,R)}{\partial R}\ .
\ee
The spatially-flat FRW universe metric is chosen as
\be
\label{FRW}
ds^2=-dt^2 + a(t)^2 \sum_{i=1}^3  \left(dx^i\right)^2\ .
\ee
Then the $(t,t)-$component of (\ref{GR2}) has the following form:
\bea
\label{GR4}
0&=& GF_G(G,R) - F(G,R) - 24 H^3
\frac{dF_G(G,R)}{dt} \nn && + 6\left(\frac{dH}{dt} +
H^2\right)F_R(G,R) - 6H\frac{dF_R(G,R)}{dt} \nn && + \rho_m\ ,
\eea
where $\rho_m$ is the energy density corresponding to matter.
Here, $G$ and $R$ have the following form:
\be
\label{GR5}
G=24\left(H^2 \dot H + H^4\right)\ ,\quad R=6\left(\dot H + 2H^2\right)\ .
\ee
In absence of matter ($\rho_m=0$), there can be a de Sitter solution
($H=H_0=$ constant) for (\ref{GR4}), in general (see \cite{BT}). One finds
$H_0$ by solving the algebraic equation
\be
\label{GR6}
0=24H_0^4 F_G(G,R) - F(G,R) + 6H_0^2 F_R(G,R)\ .
\ee
For a large number of choices of the function $F(G,R)$, Eq.~(\ref{GB6}) has a
non-trivial
($H_0\neq 0$) real solution for $H_0$ (the de Sitter universe). The 
late-time cosmology for the
above theory without matter has been discussed for a number of examples
in refs.\cite{GB}.

In this section, we  restrict the form of $F(G,R)$ to be
\be
\label{ff1}
F(G,R)=\frac{1}{2\kappa^2}R + f(G)\ ,
\ee
where $\kappa^2=8\pi G_N$, $G_N$ being the Newton constant.
As will be shown, such an action may pass the Solar System tests quite
easily. Let us consider now several different forms of such action.
By introducing two auxilliary fields, $A$ and $B$, one can rewrite
 action (\ref{GR1}) with (\ref{ff1}) as
\bea
\label{GB3}
S&=&\int d^4 x\sqrt{-g}\Bigl(\frac{1}{2\kappa^2}R + B\left(G-A\right) \nn
&& + f(A) + {\cal L}_m \Bigr)\ .
\eea
Varying over $B$, it follows that
$A=G$. Using this in (\ref{GB3}),  the action  (\ref{GR1}) with (\ref{ff1}) is
recovered. On the other hand,  varying over $A$ in (\ref{GB3}), one
gets $B=f'(A)$, and hence
\bea
\label{GB6}
S&=&\int d^4
x\sqrt{-g}\Bigl(\frac{1}{2\kappa^2}R + f'(A)G \nn && - Af'(A) +
f(A)\Bigr)\ .
\eea
By varying over $A$, the relation $A=G$ is
obtained again. The scalar is not dynamical and it has no kinetic
term. We may add, however, a kinetic term to the action by hand
\bea
\label{GB6b}
S&=&\int d^4 x\sqrt{-g}\Bigl(\frac{1}{2\kappa^2}R -
\frac{\epsilon}{2}\partial_\mu A \partial^\mu A \nn && + f'(A)G -
Af'(A) + f(A)\Bigr)\ .
\eea
Here $\epsilon$ is a positive constant parameter. 
Then, one obtains a dynamical scalar theory
coupled with the Gauss-Bonnet invariant and with a potential. It is
known that a theory of this kind  has no ghosts and it is stable, in
general. Actually, it is related with string-inspired dilaton gravity,
proposed as an alternative for dark energy \cite{string}.
Thus, in the case that the limit $\epsilon\to 0$ can be
obtained smoothly, the corresponding $f(G)$ theory would not have a
ghost and could actually be stable. This question deserves further 
investigation.

We now consider the case $\rho_m\neq 0$. Assuming that the EoS parameter
$w\equiv p_m/\rho_m$ for matter ($p_m$ is the
pressure of matter) is a constant, then, by using the conservation of
energy: $\dot \rho_m + 3H\left(\rho_m + p_m\right)=0$, we find
$\rho=\rho_0 a^{-3(1+w)}$. We also assume that $f(G)$ is given by
\be
\label{mGB1}
f(G)=f_0\left|G\right|^\beta\ ,
\ee
with constants
$f_0$ and $\beta$. If $\beta<1/2$, the $f(G)$ term becomes dominant, as
compared with the Einstein term, when the curvature is small. If we
neglect the contribution from the Einstein term in (\ref{GR4}) with
(\ref{ff1}),
assuming that
\be
\label{mGB2}
a=\left\{\begin{array}{ll} a_0t^{h_0},\ &\mbox{when}\ h_0>0 \\
a_0\left(t_s - t\right)^{h_0},\ &\mbox{when}\ h_0<0 \\
\end{array} \right. ,
\ee
the following solution is found
\bea
\label{mGB3}
h_0&=&\frac{4\beta}{3(1+w)}\ ,\nn
a_0&=&\Bigl[ -\frac{f_0(\beta - 1)}
{\left(h_0 - 1\right)\rho_0}\left\{24 \left|h_0^3 \left(- 1 +
h_0\right) \right|\right\}^\beta \nn
&& \times \left( h_0 - 1 +
4\beta\right)\Bigr]^{-\frac{1}{3(1+w)}}\ .
\eea
One can define the effective EoS parameter $w_{\rm eff}$ as
\be
\label{FRW3k} w_{\rm eff}=\frac{p}{\rho}= -1 - \frac{2\dot H}{3H^2}\ ,
\ee
which is less than $-1$ if $\beta<0$, and for $w>-1$ as
\be
\label{mGB3b}
w_{\rm eff}=-1 + \frac{2}{3h_0}=-1 +
\frac{1+w}{2\beta}\ ,
\ee
which is again less than $-1$ for
$\beta<0$. Thus, if $\beta<0$, we obtain an effective phantom with
negative $h_0$ even in the case when $w>-1$. In the phantom phase,
 a singularity of Big Rip type at $t=t_s$ \cite{brett} seems to appear 
(for the classification of these singularities, see \cite{tsujikawa}).
Near this sort of Big Rip singularity, however, the curvature becomes
dominant and then the Einstein term dominates, so that the
$f(G)$-term can be neglected. Therefore, the universe behaves as
$a=a_0t^{2/3(w+1)}$ and, as a consequence, the Big Rip singularity
will not eventually appear. The phantom era is transient.

A similar model has been found in \cite{ANO} by using a consistent
version \cite{NOPRD} of
$1/R$-gravity\cite{capozziello}. In general, in the case of 
${\cal F}(R)$-gravity
instabilities appear \cite{DK}. These instabilities do not show up
for the case of $f(G)$-gravity.

Note that under the assumption (\ref{mGB2}), the GB invariant $G$
and the scalar curvature $R$ behave as \bea \label{mGB4}
G&=&\frac{24h_0^3 \left(h_0 - 1\right)}{t^4}\ ,\ \mbox{or}\
\frac{24h_0^3 \left(h_0 - 1\right)} {\left(t_s -t\right)^4}\ , \nn
R&=&\frac{6h_0 \left(2h_0 - 1\right)}{t^2}\ ,\ \mbox{or}\ \frac{6h_0
\left(2h_0 - 1\right)}{\left(t_s -t\right)^2}\ . \eea As a
consequence, when the scalar curvature $R$ becomes small, that is,
when $t$ or $t_s - t$ becomes large, the GB invariant $G$ becomes
small more rapidly than $R$. And when $R$ becomes large, that is, if
$t$ or $t_s - t$ becomes small, then $G$ becomes large more rapidly than
$R$. Thus, if $f(G)$ is given by (\ref{mGB1}) with $\beta<1/2$, the
$f(G)$-term in the action (\ref{GR1}) with (\ref{ff1}) becomes more 
dominant for
small curvature than the Einstein term, but becomes less dominant in
the case of large curvature. Therefore, Eq.~(\ref{mGB3}) follows
when the curvature is small. There are, however, some exceptions to
this. As is clear from the expressions in (\ref{mGB4}), when
$h_0=-1/2$, which corresponds to $w_{\rm eff}=-7/3$, $R$ vanishes,
and when $h_0=-1$, corresponding to $w_{\rm eff}=-5/3$, $G$
vanishes. In both these cases, only one of the Einstein and $f(G)$ terms
survives.

In the case when $\beta<0$, if the curvature is large, the Einstein
term in the action (\ref{GR1}) with (\ref{ff1}) dominates and we have a
non-phantom
universe, but when the curvature is small, the $f(G)$-term dominates
and we obtain an effective phantom one. Since the universe starts
with large curvature, and the curvature becomes gradually smaller, the
transition between the non-phantom and phantom cases can naturally
occur in the present model.

The case when $0<\beta<1/2$ may be also considered. As $\beta$ is
positive, the universe does not reach here the phantom phase. When
the curvature is strong, the $f(G)$-term in the action (\ref{GR1}) with
(\ref{ff1})
can be neglected and we can work with Einstein's gravity. Then, if
$w$ is positive, the matter energy density $\rho_m$ should behave as
$\rho_m\sim t^{-2}$, but $f(G)$ goes as $f(G)\sim t^{-4\beta}$.
Then, for late times (large $t$), the $f(G)$-term may become
dominant as compared with the matter one. If we neglect the
contribution from matter, Eq.~(\ref{GR4}) with (\ref{ff1}) has a de Sitter
universe
solution where $H$, and therefore $G$, are constant. If $H=H_0$ with
constant $H_0$, Eq.~(\ref{GR4}) with (\ref{ff1}) looks as (\ref{GR6}) with
(\ref{ff1}). As a
consequence, even if we start from the deceleration phase with
$w>-1/3$, we may also reach an asymptotically de Sitter universe,
which is an accelerated universe. Correspondingly, also here there
could be a transition from acceleration to deceleration of the
universe.

Now, we consider the case when the contributions coming from the
Einstein and matter terms can be neglected. Then, Eq.~(\ref{GR4}) with
(\ref{ff1}) reduces to
\be
\label{mGB9}
0=Gf'(G) - f(G) - 24 \dot Gf''(G) H^3 \ .
\ee
If $f(G)$ behaves as (\ref{mGB1}), from assumption
(\ref{mGB2}), we obtain
\be
\label{mGB10}
0=\left(\beta -
1\right)h_0^6\left(h_0 - 1\right) \left(h_0 - 1 + 4\beta \right)\ .
\ee
As $h_0=1$ implies $G=0$, one may choose
\be
\label{mGB11}
h_0 = 1 - 4\beta\ ,
\ee
and Eq.~(\ref{FRW3k}) gives
\be
\label{mGB12}
w_{\rm eff}=-1 + \frac{2}{3(1-4\beta)}\ .
\ee
Therefore, if
$\beta>0$, the universe is accelerating ($w_{\rm eff}<-1/3$) and if
$\beta>1/4$, the universe is in a phantom phase ($w_{\rm eff}<-1$).
Thus, we are led to consider the following model:
\be
\label{mGB13}
f(G)=f_i\left|G\right|^{\beta_i} + f_l\left|G\right|^{\beta_l} \ ,
\ee
where we assume that
\be
\label{mGB14}
\beta_i>\frac{1}{2}\ ,\quad \frac{1}{2}>\beta_l>\frac{1}{4}\ .
\ee
Here, when the
curvature is large, as in the primordial universe, the first term
dominates, compared with the second one and the Einstein term, and
gives
\be
\label{mGB15}
-1>w_{\rm eff}=-1 +
\frac{2}{3(1-4\beta_i)}>-5/3\ .
\ee
On the other hand, when the
curvature is small, as is the case in the present universe, the
second term in (\ref{mGB13}) dominates, compared with the first one
and the Einstein term, and yields
\be
\label{mGB16} w_{\rm eff}=-1 +
\frac{2}{3(1-4\beta_l)}<-5/3\ .
\ee
Therefore, the theory (\ref{mGB13}) can in
fact produce a model which is able to describe both inflation and the
late-time acceleration of our universe in a unified way.

Instead of (\ref{mGB14}), one may also choose $\beta_l$ as
\be
\label{mGB17}
\frac{1}{4}>\beta_l>0\ ,
\ee
which gives
\be
\label{mGB18} -\frac{1}{3}>w_{\rm eff}>-1\ .
\ee
Then, what we obtain is effective quintessence. Moreover, by properly
adjusting
the couplings $f_i$ and $f_l$ in (\ref{mGB13}), we can obtain a
period where the Einstein term dominates and the universe is in a
deceleration phase. After that, there would come a transition from
deceleration to acceleration, where the GB term becomes the dominant
one.

One can consider the system  (\ref{mGB13}) coupled with matter as
in (\ref{GR4}) with (\ref{ff1}). To this end we just choose
\be
\label{mmG1}
\beta_i>\frac{1}{2}>\beta_l\ .
\ee
Then, when the curvature is
large, as in the primordial universe, the first term dominates, as
compared with the second one and the Einstein term. And when the
curvature is small, as in the present universe, the second term in
(\ref{mGB13}) is dominant as compared with the first and
Einstein's. Then, an effective $w_{\rm eff}$ can be obtained from
(\ref{mGB3b}). In the primordial universe,  matter could be
radiation with $w=1/3$, and hence the effective $w$ is
given by
\be
\label{mmG2}
w_{i,{\rm eff}}= -1 + \frac{2}{3\beta_i}\ ,
\ee
which can be less than $-1/3$, that is, the universe is
accelerating, when $\beta_i>1$. On the other hand, in the late-time
universe  matter could be dust with $w=0$, and then we would obtain
\be
\label{mmG3} w_{l,{\rm eff}}=-1 + \frac{1}{2\beta_l}\ ,
\ee
which is larger than $0$, if $0<\beta_l<1/2$, or less than $-1$, if
$\beta_l$ is negative. Thus,  acceleration could occur
in both the primordial and late-time universes, if
\be
\label{mmG4}
\beta_i>1\ ,\quad \beta_l<0\ .
\ee
Similarly, one can consider DE cosmology for other choices of f(G), for
instance,
$\ln G$ or other function $f$ increasing with the decrease of G
(late universe).

Let us address the issue of the correction to Newton's law. Let
$g_{(0)}$ be a solution of (\ref{GR2}) with (\ref{ff1}) and represent the
perturbation of the metric as $g_{\mu\nu}=g_{(0)\mu\nu} +
h_{\mu\nu}$. First, we consider the perturbation around the de Sitter
background which is a solution of (\ref{GR6}) with (\ref{ff1}). We write  the
de Sitter space metric as $g_{(0)\mu\nu}$, which gives the following
Riemann tensor:
\be
\label{GB35}
R_{(0)\mu\nu\rho\sigma}=H_0^2\left(g_{(0)\mu\rho}g_{(0)\nu\sigma}
    - g_{(0)\mu\sigma}g_{(0)\nu\rho}\right)\ .
\ee
The flat background corresponds to the limit of $H_0\to 0$. For
simplicity,  the following gauge condition is chosen:
$g_{(0)}^{\mu\nu} h_{\mu\nu}=\nabla_{(0)}^\mu h_{\mu\nu}=0$. Then
Eq.~(\ref{GR2}) with (\ref{ff1}) gives
\be
\label{GB38}
0=\frac{1}{4\kappa^2} \left(\nabla^2 h_{\mu\nu}
    - 2H_0^2 h_{\mu\nu}\right) + T_{\mu\nu}\ .
\ee
The GB term contribution  does not appear except
in the length parameter $1/H_0$ of the de Sitter space, which is
determined with account to the GB term.
This may occur due to the special structure of the GB invariant.
Eq.~(\ref{GB38}) tells us
that there is no correction to Newton's law in de Sitter and even in
the flat background corresponding to $H_0\to 0$, whatever is the
form of $f$ (at least, with the above gauge condition). (Note that
a study of the Newtonian limit in $1/R$ gravity (where significant
corrections to Newton's law may appear), and its extension
has been done in \cite{newton,NOPRD}.) For most $1/R$ models
the corrections to Newton's law do not comply with solar system tests.

Expression (\ref{GB38}) can be actually valid
in the de Sitter background only. In a more general FRW universe, there
can appear corrections coming from the $f(G)$ term. We should also
note that in deriving (\ref{GB38}),  a gauge condition
$g_{(0)}^{\mu\nu} h_{\mu\nu}=0$ was used, but if  the mode
corresponding to $g_{(0)}^{\mu\nu} h_{\mu\nu}$ is included, there might
appear corrections coming from the $f(G)$ term. The mode corresponding to
$g_{(0)}^{\mu\nu} h_{\mu\nu}$ gives an infinitesimal scale
transformation of the metric. Then, it is convenient to write the metric as
\be
\label{ST1}
g_{\mu\nu}=\e^\phi g_{(0)\mu\nu}\ .
\ee
Here
$g_{(0)\mu\nu}$ expresses the metric of de Sitter space in
(\ref{GB35}). The Gauss-Bonnet invariant $G$ is correspondingly
given by
\bea
\label{ST2}
G&=&\e^{-2\phi}\left\{24H_0^4 - 12 H_0^2
\nabla_{(0)}^2\phi
    - 6H_0^2\partial_\mu\phi\partial^\mu\phi \right. \nn
&& + 2 \left(\nabla_{(0)}^2 \phi\right)^2
    - 2 \nabla_{(0)\mu} \nabla_{(0)\nu}\phi\nabla_{(0)}^\mu \nabla_{(0)}^\nu\phi
\nn
&& \left. + \nabla_{(0)}^2 \partial_\mu\phi\partial^\mu\phi
+ 2 \nabla_{(0)\mu} \nabla_{(0)\nu}\phi\partial_{(0)}^\mu\phi
\partial_{(0)}^\nu\phi \right\} \nn
&=& \e^{-2\phi}\left\{24H_0^4 + \nabla_{(0)}^\mu \left( - 12 H_0^2
\nabla_{(0)\mu}\phi \right.\right. \nn
&& + 2 \partial_\mu \phi \nabla_{(0)}^2 \phi
    - 2 \nabla_{(0)\mu} \nabla_{(0)\nu}\phi \nabla_{(0)}^\nu\phi \nn
&& \left.\left. + \partial_\mu\phi\partial_{(0)}^\nu\phi
\partial_{(0)\nu}\phi\right) \right\} \ .
\eea
The
covariant derivative associated with $g_{(0)\mu\nu}$ is written here as
$\nabla_{(0)\mu}$. The expansion of $f(G)$, with respect to $\phi$, is
\bea
\label{ST3}
&& \sqrt{-g}f(G) \nn
&& =\sqrt{-g_{(0)}}\left\{f\left(24H_0^4\right) + 2
f\left(24H_0^4\right) \phi^2 \right. \nn
&& - 48 f'\left(24H_0^4\right) H_0^4 \phi^2 \nn
&& + 72 f''\left(24H_0^4\right)H_0^4\left(\nabla_{(0)}^2 \phi +
4H_0^2\phi\right)^2 \nn && \left. + {\cal O}\left(\phi^3\right) +
\mbox{total derivative terms} \right\}\ .
\eea
Since the last term
contains $\left(\nabla_{(0)}^2 \phi \right)^2$, in general, there could
 be an instability. A way to avoid the problem is to
fine-tune $f(G)$ so that $f''\left(24H_0^4\right)$  vanishes for
the solution $H_0$ in (\ref{GR6}) with (\ref{ff1}).

In order to consider a more general case, one expands $f(G)$ in the
action (\ref{GR1}) with (\ref{ff1}) as
\bea
\label{GBp1}
f(G) &=& f\left(G_{(0)}\right) + \frac{1}{2}f''\left(G_{(0)}\right) \delta G^2
\nn
&& + f'\left(G_{(0)}\right) \delta^2 G + {\cal O}\left(h^3\right)\ ,
\eea
where $G_{(0)}$ is the Gauss-Bonnet
invariant given by $g_{(0)\mu\nu}$ and
\bea
\label{GBp2}
\delta G &=& 4 \left(R \nabla^\mu \nabla^\nu h_{\nu\mu}
    - 4 R^\rho_{\ \nu}R^{\nu\sigma} h_{\rho\sigma} \right. \nn
&& + 4 R^{\nu\sigma}R^{\rho\ \mu}_{\ \sigma\ \nu} h_{\mu\rho} \nn
&& \left. -4 \nabla_\nu \nabla^\mu h_{\sigma\mu}
    + R^{\mu\nu\rho\sigma} \nabla_\rho \nabla_\nu h_{\sigma\mu}\right)\ ,\\
\label{GBp3}
\delta^2 G &=& \left[h_{\zeta\eta}\left\{ - R\left(\nabla^\eta \nabla^\mu
h_\mu^{\ \zeta}
+ \nabla^\mu \nabla^\eta h_\mu^{\ \zeta} \right) \right. \right. \nn
&& - R^{\zeta\eta}\nabla^\rho \nabla^\nu h_{\nu\rho}
+ 4 R^{\nu\sigma}\nabla^\eta \nabla_\nu h_\sigma^{\ \zeta} \nn
&& + 2 R^{\zeta\sigma} \nabla^\mu\nabla^\eta h_{\sigma\mu}
+ 4 R^{\nu\zeta}\nabla^\mu \nabla_\nu h_\mu^{\ \eta} \nn
&& + 2 R^{\zeta\nu\eta\sigma}\nabla^\mu\nabla_\nu h_{\sigma\mu}
    - R^{\mu\nu\zeta\sigma}\nabla^\eta \nabla_\nu h_{\sigma\mu} \nn
&& - \frac{1}{2}R^{\zeta\nu\rho\sigma}\left( \nabla_\rho \nabla_\nu
h_\sigma^{\ \eta}
    - \nabla_\rho \nabla^\eta h_{\sigma\nu}\right) \nn
&& \left. - R^{\mu\nu\rho\zeta} \nabla_\rho \nabla_\nu h^\eta_{\ \mu}\right\}
\nn
&& + \frac{1}{2}\left\{ 2\nabla^\mu \nabla^\nu h_{\nu\mu}\left(\nabla^\alpha
\nabla^\beta h_{\alpha\beta}
    - \nabla^2 h_\alpha^\alpha \right) \right. \nn
&& -4\nabla^\mu \nabla_\nu h_{\sigma\mu} \left(\nabla^\alpha \nabla^\nu
h_\alpha^{\ \sigma}
    - \nabla^2 h^{\sigma\nu} \right. \nn
&& \left. - \nabla^\sigma \nabla^ \nu h_\alpha^{\ \alpha}
+ \nabla^\sigma \nabla^\alpha h_\alpha^{\ \nu}\right) \nn
&& + \nabla_\rho \nabla_\nu h_{\sigma\mu} \left( \nabla^\rho \nabla^\nu
h^{\sigma\mu}
- \nabla^\rho \nabla^\mu h^{\sigma\nu} \right. \nn
&& \left. \left. - \nabla^\sigma \nabla^\nu h^{\rho\mu}
+ \nabla^\sigma \nabla^\mu h^{\sigma\nu} \right)\right\} \nn
&& - \frac{1}{2}\left\{ R \left(2\nabla^\mu h_{\nu\lambda} \nabla^\lambda
h^\nu_{\ \mu}
+ 2\nabla^\mu h_{\mu\nu} \nabla_\nu h^{\nu\lambda} \right. \right. \nn
&& - \nabla_\lambda h_\alpha^{\ \alpha}\nabla_\nu h^{\nu\lambda}
+ \nabla_\mu \left( 2 h^{\lambda\mu}\nabla^\nu h_{\nu\lambda} \right. \nn
&& \left. \left. - h^{\lambda\mu}\nabla_\lambda h_\alpha^{\ \alpha}
+ h^{\nu\lambda} \nabla^\mu h_{\nu\lambda} \right) \right) \nn
&& -4 R^{\nu\sigma}\left\{\left(\nabla^\mu h_{\nu\lambda} + \nabla_\nu
h^\mu_{\ \lambda} \right. \right. \nn
&& \left. - \nabla_\lambda h_\nu^{\ \mu} \right) \nabla^\lambda h_{\sigma\mu}
\nn
&& + \left( \nabla_\sigma h^\mu_{\ \lambda} + \nabla^\mu h_{\sigma\lambda}
    - \nabla_\lambda h^\mu_{\ \sigma}\right)\nabla_\nu h^\lambda_{\ \mu} \nn
&& + \left(2\nabla^\mu h_{\mu\lambda} - \nabla_\lambda h_\alpha^{\ \alpha}
\right)
\nabla_\nu h_\sigma^{\ \lambda} \nn
&& + \nabla_\mu \left(2h^{\lambda\mu} \nabla_\nu h_{\sigma\lambda}\right.\nn
&& \left. \left. \left. + \left( \nabla_\nu h_{\mu\lambda}
+ \nabla_\mu h_{\nu\lambda} - \nabla_\lambda h_{\nu\mu} \right)
h_\sigma^{\ \lambda} \right) \right) \right\} \nn
&& + R^{\mu\nu\rho\sigma}\left( \left( \nabla_\rho h_{\nu\lambda} + \nabla_\nu
h_{\rho\lambda}
\right.\right. \nn
&& \left. - \nabla_\lambda h_{\nu\rho}\right) \nabla^\lambda h_{\sigma\mu} \nn
&& + \left( \nabla_\mu h_{\rho\lambda} + \nabla_\rho h_{\mu\lambda}
    - \nabla_\lambda h_{\rho\mu}\right) \nabla_\nu h_{\sigma\lambda} \nn
&& + \nabla_\rho \left( \left( \nabla_\nu h_{\sigma\lambda} +
\nabla_\sigma h_{\nu\lambda} \right.\right. \nn
&& \left. \left. \left. \left. - \nabla_\lambda h_{\nu\sigma}
\right)h^\lambda_{\
\mu} \right)\right\} \right]\ .
\eea
Note that the term proportional to $\delta G$ does not appear in
(\ref{GBp1}) since
the background metric is a solution of (\ref{GB35}).
Here and in the following
the index (0) is always suppressed, when there cannot be confusion. If we
choose the gauge condition $\nabla^\mu h_{\mu\nu}=0$,
Eqs.~(\ref{GBp2}) and (\ref{GBp3}) have the following form:
\bea
\label{GBp4}
\delta G &=& 4 \left(  - 4 R^\rho_{\ \nu}R^{\nu\sigma}
h_{\rho\sigma} + 4 R^{\nu\sigma}R^{\rho\ \mu}_{\ \sigma\ \nu}
h_{\mu\rho}  \right. \nn
&& \left. + R^{\mu\nu\rho\sigma} \nabla_\rho \nabla_\nu h_{\sigma\mu}\right)\
,\\
\label{GBp5}
\delta^2 G &=& \left[h_{\zeta\eta}\left\{ - R\left(
h^{\rho\zeta}R_\rho^{\ \eta} - h_{\mu\rho}R^{\rho\zeta\mu\eta}\right) \right.
\right. \nn
&& + 4 R^{\nu\sigma}\nabla^\eta \nabla_\nu h_\sigma^{\ \zeta} \nn
&& + 2 R^{\zeta\sigma}
\left( h_{\rho\sigma}R^{\rho\eta} - h_{\mu\rho} R^{\rho\ \mu\eta}_{\
\sigma}\right) \nn
&& + 4 R^{\nu\zeta}
\left(h^{\rho\eta}R_{\rho\nu} - h_{\mu\rho} R^{\rho\eta\mu}_{\ \ \ \
\nu}\right) \nn
&& + 2 R^{\zeta\nu\eta\sigma}
\left(h_{\rho\sigma}R^\rho_{\ \nu} - h_{\mu\rho} R^{\rho\ \mu}_{\ \sigma\
\nu}\right) \nn
&& - R^{\mu\nu\zeta\sigma}\nabla^\eta \nabla_\nu h_{\sigma\mu} \nn
&& - \frac{1}{2}R^{\zeta\nu\rho\sigma}\left( \nabla_\rho \nabla_\nu
h_\sigma^{\ \eta}
    - \nabla_\rho \nabla^\eta h_{\sigma\nu}\right) \nn
&& \left. - R^{\mu\nu\rho\zeta} \nabla_\rho \nabla_\nu h^\eta_{\ \mu}\right\}
\nn
&& + \frac{1}{2}\left\{ -4\left(h_{\rho\sigma}R^\rho_{\ \nu} - h_{\mu\rho}
R^{\rho\ \mu}_{\ \sigma\ \nu}\right)
\left( h_\rho^{\ \sigma}R^{\rho\nu} \right. \right. \nn
&& \left. - h_{\mu\rho} R^{\rho\sigma\mu\nu}
    - \nabla^2 h^{\sigma\nu} - \nabla^\sigma \nabla^ \nu h_\alpha^{\ \alpha}
\right) \nn
&& + \nabla_\rho \nabla_\nu h_{\sigma\mu} \left( \nabla^\rho \nabla^\nu
h^{\sigma\mu}
- \nabla^\rho \nabla^\mu h^{\sigma\nu} \right. \nn
&& \left. \left. - \nabla^\sigma \nabla^\nu h^{\rho\mu}
+ \nabla^\sigma \nabla^\mu h^{\sigma\nu} \right)\right\} \nn
&& - \frac{1}{2}\left\{ R \left(2\nabla^\mu h_{\nu\lambda} \nabla^\lambda
h^\nu_{\ \mu}
    - \nabla_\lambda h_\alpha^{\ \alpha} \right. \right. \nn
&& \left. + \nabla_\mu \left( -h^{\lambda\mu}\nabla_\lambda h_\alpha^{\
\alpha}
+ h^{\nu\lambda} \nabla^\mu h_{\nu\lambda} \right) \right) \nn
&& -4 R^{\nu\sigma}\left\{\left(\nabla^\mu h_{\nu\lambda} + \nabla_\nu
h^\mu_{\ \lambda} \right. \right. \nn
&& \left. - \nabla_\lambda h_\nu^{\ \mu} \right) \nabla^\lambda h_{\sigma\mu}
\nn
&& + \left( \nabla_\sigma h^\mu_{\ \lambda} + \nabla^\mu h_{\sigma\lambda}
    - \nabla_\lambda h^\mu_{\ \sigma}\right)\nabla_\nu h^\lambda_{\ \mu} \nn
&& - \nabla_\lambda h_\alpha^{\ \alpha} \nabla_\nu h_\sigma^{\ \lambda}
+ \nabla_\mu \left(2h^{\lambda\mu} \nabla_\nu h_{\sigma\lambda}\right.\nn
&& \left. \left. \left. + \left( \nabla_\nu h_{\mu\lambda}
+ \nabla_\mu h_{\nu\lambda} - \nabla_\lambda h_{\nu\mu} \right)
h_\sigma^{\ \lambda} \right) \right) \right\} \nn
&& + R^{\mu\nu\rho\sigma}\left( \left( \nabla_\rho h_{\nu\lambda} + \nabla_\nu
h_{\rho\lambda}
\right.\right. \nn
&& \left. - \nabla_\lambda h_{\nu\rho}\right) \nabla^\lambda h_{\sigma\mu} \nn
&& + \left( \nabla_\mu h_{\rho\lambda} + \nabla_\rho h_{\mu\lambda}
    - \nabla_\lambda h_{\rho\mu}\right) \nabla_\nu h_{\sigma\lambda} \nn
&& + \nabla_\rho \left( \left( \nabla_\nu h_{\sigma\lambda} +
\nabla_\sigma h_{\nu\lambda} \right.\right. \nn
&& \left. \left. \left. \left. - \nabla_\lambda h_{\nu\sigma}
\right)h^\lambda_{\
\mu} \right)\right\} \right]\ .
\eea
Now, we consider the case that
$\dot H \sim H^2$ in the FRW universe (\ref{FRW}). Then, by
specifying the dimension,  the following structure is found
\bea
\label{GBA4}
\delta G &\sim& H^4 h + H^2 \nabla^2 h\ , \nn
\delta^2 G &\sim& H^4 h^2 + H^2 h\nabla^2 h + H^2 \left(\nabla h\right)^2 \nn
&& + H^3 h \nabla h + \left(\nabla^2 h\right)^2\ .
\eea
For
the qualitative arguments that follow, we have abbreviated the
vector indices and coefficients. Since Eq.~(\ref{GBp1}) contains
$\delta G^2$ and $\delta^2 G$ terms, the $H^2 \nabla^2 h$-term in
$\delta G$ and $\left(\nabla^2 h\right)^2$-term in $\delta^2 G$ have
a possibility to generate the instability. Explicit calculations in
the FRW universe tell us that the $\left(\nabla^2 h\right)^2$ term
in $\delta^2 G$ vanishes identically, while the $H^2 \nabla^2 h$
term in $\delta G$ has the following form:
\be
\label{GBA6}
\mbox{$H^2 \nabla^2 h$ term in $\delta G$} = - 4 \dot H \nabla^2 h_{tt}\ .
\ee
For simplicity, we have chosen again the gauge condition
$g^{\mu\nu}_{(0)} h_{\mu\nu}=\nabla^\mu h_{\mu\nu}=0$. Then,
except for the $\dot H=0$ case, which describes the de Sitter
universe, there might be an instability.

Since the $\delta G^2$-term has a factor $f''(G_{(0)})$, if one
properly chooses the form of $f(G)$ and fine-tunse the coefficients,
it could occur that $f''\left(G_{(0)}\right)=0$ in the present universe.
Correspondingly, the term $\left(\nabla^2 h\right)^2$ does not appear
 in the action and no instability appears.

To summarize, in both cases (\ref{ST3}) and (\ref{GBA4}), if we
choose $f''=0$ in the present universe,  the instability does {\it
not} appear. As an example, one can consider the model (\ref{mGB13}).
As
\be
\label{GBex1}
f''(G)=f_i\beta_i\left(\beta_i - 1\right)
\left|G\right|^{\beta_i-2} + f_l\beta_l\left(\beta_l - 1\right)
\left|G\right|^{\beta_l-2} \ ,
\ee
if we choose the parameters $f_i$, $\beta_i$, $f_l$, and $\beta_l$ to satisfy
\be
\label{GBex2}
0=f_i\beta_i\left(\beta_i - 1\right)
\left|G_{(0)}\right|^{\beta_i-\beta_l} + f_l\beta_l\left(\beta_l - 1\right) \
,
\ee
we find $f'\left(G_{(0)}\right)=0$ and thus there
will be no instability. In (\ref{GBex2}), $G_{(0)}$ is the value of
the Gauss-Bonnet invariant given by the curvature in the present
universe.

One now rewrites (\ref{GR2}) with (\ref{ff1}) as an FRW equation:
\bea
\label{GBfrw1}
0 &=& -\frac{3}{\kappa^2}H^2 + \rho_G + \rho_M \ ,\\
0 &=& \frac{1}{\kappa^2}\left(2\dot H + 3H^2\right) + p_G + p_m \ .
\eea
Here $\rho_G$ and $p_G$ express the contribution from the
$f(G)$ term in the action (\ref{GR1}) with (\ref{ff1}):
\bea
\label{GBfrw2}
\rho_G &=& Gf'(G) - f(G) - 24 \dot G f'' (G) H^3 \ ,\nn
p_G &=& -  Gf'(G) + f(G) + 24 \dot G f'' (G) H^3 \nn && + 8\dot G^2 f'''(G)H^2
\nn
&& - 192 f''(G)\left( - 8 H^3 \dot H \ddot H - 6 H^2 \dot H^3 - H^4
\dddot H \right. \nn
&& \left. - 3H^5 \ddot H - 18 H^4 \dot H^2 + 4
H^6 \dot H\right)\ .
\eea
One can view the contribution from the
$f(G)$ term as a sort of matter satisfying a special (inhomogeneous)
EoS \cite{inh} (or usual EoS with time-dependent bulk viscosity
\cite{barrow}) of the form
\bea
\label{GBfrw3}
0&=& \rho_G + p_G
    - 8\dot G^2 f'''(G)H^2 \nn && + 192 f''(G)\left( - 8 H^3 \dot H
\ddot H - 6 H^2 \dot H^3 - H^4 \dddot H \right. \nn
&& \left. - 3H^5 \ddot H - 18 H^4 \dot H^2 + 4 H^6 \dot H\right)\ .
\eea
In particular, in the case of de Sitter space, owing to the fact
that the Hubble rate $H$, and therefore $G$, are constant, we find
$0 = \rho_G + p_G$. In the case that $f(G)$ is given by (\ref{mGB1}) and
one further assumes (\ref{mGB2}), one gets
\bea
\label{GBfrw4}
\rho_G&=& f_0\left|G\right|^\beta \left(\beta - 1\right) \frac{h_0 -
1 + 4\beta}{h_0 - 1}\ , \nn p_G&=& f_0\left|G\right|^\beta
\left(\beta - 1\right) \nn && \times \frac{3h_0^2 - (3+8\beta)h_0 +
16\beta^2 - 4\beta}{3h_0\left(h_0-1\right)}\ .
\eea
It follows that
the effective EoS $w_G\equiv p_G/\rho_G$ for the
$f(G)$ part is given by
\be
\label{GBfrw5} w_G=\frac{3h_0^2 -
(3+8\beta)h_0 + 16\beta^2 - 4\beta}{3h_0\left(h_0-1 +
4\beta\right)}\ .
\ee

In absence of matter ($\rho_m=p_m=0$), Eq.~(\ref{GBfrw1}) may be
rewritten as \be \label{GBfrw6} H^2=\frac{\kappa^2}{6}\rho_G\ ,\quad
\dot H =
    - \frac{\kappa^2}{2}\left(\rho_G + p_G\right)\ .
\ee
Then, by using the expression (\ref{GR5}), we find
\bea
\label{GBfrw7}
G&=&-\frac{2\kappa^4}{3}\rho_G\left(2\rho_G + 3p_G\right)\ ,\\
\dot G&=&4\left\{\kappa^2 \ddot H \rho + \kappa^4 \left(\rho_G +
p_G\right)\left(\rho_G + 3p_G\right)
\sqrt{\frac{\kappa^2\rho}{6}}\right\}\ , \nonumber
\eea
and by using the first equation in (\ref{GBfrw2}),  the effective equation
of state is:
\bea
\label{GBfrw8}
&& 0= - \rho_G  \nn
&& - \frac{2\kappa^4}{3}\rho_G\left(2\rho_G + 3p_G\right) f'\left(-
\frac{2\kappa^4}{3}\rho_G\left(2\rho_G + 3p_G\right)\right) \nn
&& - f\left(- \frac{2\kappa^4}{3}\rho_G\left(2\rho_G + 3p_G\right)\right) \nn
&& - (24)^2\left\{\ddot H \left(\frac{\kappa^2}{6}\rho_G\right)^{5/2} \right.
\nn
&& \left. + \frac{\kappa^2}{6}\left(\rho_G + p_G\right)\left(\rho_G +
3p_G\right) \left(\frac{\kappa^2}{6}\rho_G\right)^2 \right\} \nn
&& \times f''\left( - \frac{2\kappa^4}{3}\rho_G\left(2\rho_G +
3p_G\right) \right)\ ,
\eea
which has the form of $\tilde F\left(\rho_G,p_G, \ddot H\right)=0$ \cite{inh}.
At the dynamical level, this demonstrates the equivalency between 
modified GB gravity and the effective inhomogeneous EoS description.
It may be of interest to study the choice of $f(G)$ which leads to the
inhomogeneous generalization of EoS $p=-\rho-A\rho^\alpha$ suggested and
studied in \cite{st} as this may be easily compared with standard $\Lambda$CDM
cosmology.

Summing up, late-time cosmology in modified GB gravity with matter was
studied. It has been shown that effective DE of quintessence, phantom or
cosmological constant type can be actually produced within such theory,
with the possibility to unify it also with primordial inflation. Moreover,
the transition from the deceleration to the acceleration era  easily
occurs in some versions of the $f(G)$ theory which comply with 
the solar system tests (no instabilities appear, no corrections to 
Newton's law follow).


\section{De Sitter solution in Modified Gauss-Bonnet gravity models}

In this section, we will investigate the properties of some black hole
solutions within the modified GB gravity scheme. The role played by static,
spherically symmetric solutions, as the Schwarzschild one,
 with regard to solar system tests, is well known. Due to the absence of 
this kind of zero-curvature BH in the $f(G)$ theory
(see the Appendix) deSitter space as well as SdS BH need her special 
consideration, what it is done below.


For the general model, it can be shown that the equations of motion have 
the following structure  (no-matter case)
\beq
0 &=&\frac{1}{2}\hat F(G,R)g_{\mu\nu}-F_R(G,R)R_{\mu\nu}
\nonumber  \\
&-&2F_G(G,R)\at RR_{\mu\nu}-2R_{\mu\rho}R_{\nu}^{\rho}
+R_{\mu \rho \alpha \beta}
R_{\nu}^{\,\,\rho \alpha \beta}\cp \nonumber \\
&-&\ap 2R_{\mu \alpha \nu \beta}
R^{\alpha \beta}\ct
+\Xi_{\mu \nu}(\nabla_\alpha F_G(G,R), \nabla_\alpha F_R(G,R))\,,
\nonumber
\label{EQ}
\eeq
where $\Xi_{\mu \nu}$ is vanishing when $G$ and $R$ are constant.

In this section we are interested in finding condition assuring
 the existence of solutions of the de Sitter type (including SdS BH).
In such case, the Gauss-Bonnet invariant and the Ricci scalar are constant
and the Gauss-Bonnet invariant reads
\beq
R=R_0\,\hs G=G_0=\frac{1}{6}R_0^2\,.
\eeq
Assuming the maximally symmetric metric solution, one gets
\beq
2 F_R(G,R)R^0_{\mu\nu}= \at
 F(G_0,R_0)-G_0  F_{G_0}(G_0,R_0)\ct g^0_{\mu\nu} \,.
\label{EQ4}\eeq
Note that if $ F(G,R)=R+\alpha G$, namely there is only  a linear term in
$G$,
one gets the ordinary
Einstein equation in vacuum, as it should be, because in this case, 
$F(G,R)$ contains the Hilbert-Einstein term plus  a topological invariant.

Taking the trace of (\ref{EQ4}),  the condition follows
\beq
\frac{R_0 F_{R_0}(G_0,R_0)}{2}=\aq F(G_0,R_0)-G_0 F_{G_0}(G_0,R_0)\cq\,.
\label{exG}
\eeq
This condition will play an important role in the following and it is
equivalent to the condition (\ref{GR6}).
As an example, when $F(G,R)=R+f(G)$, (here $2 \kappa^2=1$), one has
\beq
G_0 f'(G_0)- f(G_0)=\frac{R_0}{2}\,.
\label{exG1}
\eeq
In general, solving Eq. (\ref{exG}) in terms of $R_0$, one can rewrite the
maximally symmetric  solution as
\beq
R^0_{\mu\nu}=\frac{R_0}{4}g^0_{\mu\nu}=\Lambda_{eff}\,g^0_{\mu\nu}\,,
\eeq
which defines an effective cosmological constant. For example,
when $f(G,R)=R+f(G)$, one has
\beq
\Lambda_{eff}=\frac{1}{2} \at G_0  f'(G_0)-  f(G_0)\ct\,.
\label{exG2}
\eeq
Thus, if $f(G)=-\alpha  G^{\beta}$, one has,
\beq
2\alpha(1- \beta)  \at \frac{1}{6}\ct^{\beta}=R_0^{1-2\beta}\,.
\eeq
When $\beta$ is small, $\alpha >0$, one obtains $R_0 \sim \alpha$, while
with the choice $\beta=-1/2$, $\alpha >0$, one has
\beq
R_0=6^{1/4}\sqrt{3 \alpha}\,,
\label{invG}
\eeq
and the corresponding effective cosmological constant reads
\beq
\Lambda_{eff}=\frac{1}{4}
6^{1/4}\sqrt{3 \alpha}\,.
\eeq

As in the pure Einstein case, one is confronted with  the black hole
nucleation problem \cite{ginspar}. We review here the  discussion
reported in refs. \cite{ginspar,wipf}.

To begin with, we recall that we shall deal with a tunneling process in
quantum gravity. On general backgrounds, this process is mediated by the
associated gravitational instantons, namely stationary solutions of Euclidean
gravitational action, which dominate the path integral of Euclidean quantum
gravity. It is a well known fact that as soon as an imaginary part appears in
the one-loop partition
function, one has a metastable thermal state and thus
a non vanishing decay rate.
Typically, this imaginary part comes from the existence of a negative mode
in the one-loop functional determinant.
Here, the semiclassical and one-loop approximations are
the only techniques at disposal, even though one should bear in mind their
limitations as well as their merits.

Let us consider  a general model described
by $F(G, R)$, satisfying the  condition (\ref{exG}) and  with $\Lambda_{eff} >0$.
Thus, we may have a de Sitter  Euclidean
instanton. In the Euclidean version, the associated manifold is $S_4$.

Making use of the  instanton approach,
for the Euclidean partition function we have
\beq
Z&\simeq& Z(S_4)
\nonumber\\
&=&Z^{(1)}(S_4)e^{-I(S_4)}\,,\eeq
where $I$ is the
classical action and $Z^{(1)}$  the quantum correction, typically a
ratio of functional determinants. The classical action can be easily
evaluated and reads
\beq
I(S_4)=-\frac{192 \pi^2}{\kappa^2 R^2_0} F(G_0,R_0) \,.
\label{a}
\eeq

At this point we make a brief  disgression regarding the entropy of
the above black hole solution. To this aim, we follow the arguments
reported in Ref.~\cite{brevik}. If one make use of the Noether
charge method \cite{wald} for evaluating the entropy associated with black hole
solutions with constant Gauss-Bonnet and Ricci invariants in modified gravity
models,  a direct computation gives
\be S=\frac{2\pi A_H}{\kappa^2}\at F_{R_0}(G_0,R_0)
+\frac{ R_0 }{3} F_{G_0}(G_0,R_0)\ct \,.
\label{ve}
\ee
In the
above equations, $A_H=4\pi r_H^2$,  $r_H$ being the radius of the
event horizon or cosmological horizon related to a black hole
solution. This turns out to be model dependent.

Another consequence, as is well known, is the modification of the
``Area Law'', which reads instead
\be S=\frac{2\pi A_H}{\kappa^2}=\frac{ A_H}{4 G_N} \,.
\label{al}
\ee
One should also stress that, since the above entropy formula
depends on $F_{G}(G,R)$, there is always an indetermination: any
 linear term
in $G$ appearing in the classical action is irrelevant as far as the
equations of motion are concerned, while in the entropy formula it
gives a constant non vanishing contribution. This is a kind of
indetermination associated with the Noether method \cite{wald}.

Furthermore, as in principle the quantity which modifies in a non
trivial way the usual Area Law
\beq
F_{R_0}(G_0,R_0)
+\frac{ R_0 }{3} F_{G_0}(G_0,R_0)
\eeq
might be negative, there exists the possibilty of having negative BH
entropies, according to specific choices of $F(G,R)$.

Let us consider an example. For the model defined by $F(G,R)=R+f(G)$,
\beq
1+\frac{ R_0}{3} f'(G_0)=2 \at 1+\frac{f(G_0)}{R_0} \ct\,.
\eeq
Thus, with the choice $f(G)=-\alpha G^\beta$, $\alpha >0$, one has
\beq
S=\frac{4\pi A_H}{ \kappa^2}\frac{1}{\beta-1}\,.
\label{ve1}
\eeq
As a result, modulo the Noether charge method indetermination, the
entropy may be negative.

Coming back to the general case, we recall that we are interested
in the  de Sitter metric, which reads
\beq
ds^2=-\at 1-\frac{r^2}{l^2}\ct dt^2 +
 \frac{dr^2}{1-\frac{r^2}{l^2}}+r^2dS^2_2\,,
\eeq
with $\Lambda_{eff}=\frac{3}{l^2}$. The Ricci scalar is $R_0=4 \Lambda$.

Since $r_H=l$, one has
\be A_H=\frac{12\pi}{\Lambda_{eff}}=\frac{48 \pi}{R_0}\,. \ee
As a consequence,
\be S(S_4)=\frac{96  \pi^2}{\kappa^2  R_0}
\at F_{R_0}(G_0,R_0)
+\frac{ R_0 }{3} F_{G_0}(G_0,R_0)\ct \,.
\label{ve2} \ee
Taking Eqs. (\ref{exG}) and (\ref{a}) into account, from the last equation, one obtains
\be S(S_4)=-I(S_4)\,,
\ee
which is a good check of our entropy formula (\ref{ve}).

\section{One-loop quantization of modified Gauss-Bonnet gravity on de
Sitter space}

Here we discuss the one-loop quantization of the class of models we
are dealing with, on a maximally symmetric space.
One-loop contributions are certainly important during the inflationary
phase, but as it has been shown in \cite{Cognola:2005sg}, they also
provide a powerful method in order to study the stability of the
solutions.

We start by
recalling some properties of the classical model defined by the
the choice $F(G,R)=\frac{1}{2\kappa^2}\at R+ \hat f(G) \ct $, where now the generic function $\hat f(G)$
is supposed to satisfy the ``on shell'' condition (\ref{exG1}). Such condition
ensures the existence of a constant GB invariant,
 maximally symmetric solution
of the field equations (\ref{GR2}) with (\ref{ff1}).

In accordance with the background field method, we now consider the
small fluctuations of the fields around the de Sitter manifold. Then,
for the arbitrary solutions of the field equations we set
\beq
g_{\mu\nu}=g_{0\mu\nu}+h_{\mu\nu} \eeq
and perform a Taylor expansion of the action around the de Sitter
manifold. Up to second order in $h_{\mu\nu}$, we get
\beq
S[h]&\sim& \frac{1}{2\ka^2}\int\:d^4x\,\sqrt{-g_0}\:
     \aq R_{0}+\hat f_0\cp
\nonumber\\ &&\ap
+\at\frac{R_0}4+\frac{\hat f_0}2
     -\frac{R_{0}^2\hat f_1}2\ct\,h +{\cal L}_2\cq\,,
\label{AAA3} \eeq
where ${\cal L}_2$ represents the quadratic
contribution in the fluctuation field $h_{\mu\nu}$ and,
in contrast with previous sections,
here $\nabla_\mu$ represents the covariant derivative
in the unperturbed metric $g_{0\mu\nu}$.
For the sake of simplicity, we have also used the notation
$\hat f_0=\hat
f(G_{0})$, $\hat f_1=\hat f'(G_{0})$ and $\hat f_2=\hat f''(G_{0})$.


For technical reasons it is convenient to
carry out the standard expansion of the tensor field $h_{\mu\nu}$ in
irreducible components \cite{frad}, that is
\beq h_{\mu\nu}&=&\hat
h_{\mu\nu}+\nabla_\mu\xi_\nu+\nabla_\nu\xi_\mu\nonumber\\ && \hs
+\nabla_\mu\nabla_\nu\sigma +\frac14\,g_{\mu\nu}(h-\lap\sigma)\:,
\label{tt}\eeq
where $\si$ is the scalar component, while $\xi_\mu$
and $\hat h_{\mu\nu}$ are the vector and tensor components, with the
properties
\beq \nabla_\mu\xi^\mu=0\:,\hs \nabla_\mu\hat h^\mu_\nu=0\:,\hs
\hat h^\mu_\mu=0\:. \label{AAA4} \eeq
In terms of the irreducible
components of the $h_{\mu\nu}$ field, the quadratic part of the
Lagrangian density, disregarding total derivatives, reads
\begin{eqnarray}
{\cal L}_2 = {\cal L}_{tensor}+{\cal L}_{vector}+{\cal L}_{scalar}\,,
\end{eqnarray}
where ${\cal L}_{tensor}$, ${\cal L}_{vector}$ and ${\cal L}_{scalar}$
represent the tensor, vector and scalar contributions
respectively. They have the form \beq {\cal L}_{tensor} &=& \hat
h_{\mu\nu}\,\aq{-\frac{{\hat f_{0}}}{4}} -{\frac{R_0}{6}} +{\frac{{\hat
f_{1}}\,{R_0^2}}{6}} \cp\nonumber \\ &&\ap +\at\frac14+\frac{\hat
f_1\,R_0}6\ct\Delta_2\,\cq\,\hat h^{\mu\nu}\,, \eeq
\beq {\cal L}_{vector}&=& \xi_\mu\,\aq{\frac{{\hat
f_{0}}\,R_0}{8}}+{\frac{{R_0^2}}{16}} -{\frac{31\,{\hat
f_{1}}\,{R_0^3}}{288}} -{\frac{{\hat f_{2}}\,{R_0^5}}{432}}
\cp\nonumber\\ && +\at{\frac{{\hat
f_{0}}}{2}}+{\frac{R_0}{4}}-{\frac{17\,{\hat f_{1}}\,{R_0^2}}{36}}
+{\frac{{\hat f_{2}}\,{R_0^4}}{108}}\ct\,\Delta_1 \nonumber\\ &&
\ap\hs\hs\hs -\frac{\hat f_1\,R_0}{32}\Delta_1^2\cq\,\xi^\mu\,,
\eeq
\beq {\cal L}_{scalar}&=& h\,\aq{\frac{{\hat f_{0}}}{16}}
+{\frac{{\hat f_{1}}\,{R_0^2}}{48}} +{\frac{{\hat
f_{2}}\,{R_0^4}}{72}} \cp\nonumber\\ &&\hs
+\at-\frac{3}{32}+{\frac{5\,{\hat f_{1}}\,R_0}{32}} +{\frac{{\hat
f_{2}}\,{R_0^3}}{24}}\ct\,\Delta_0 \nonumber\\ &&\ap\hs\hs\hs
+\frac{{\hat f_{2}}\,{R_0^2}}{32}\,\Delta_0^2\cq\,h \nonumber\\&&
+\sigma\,\aq {-\frac{{\hat f_{0}}\,R_0}{16}}
-{\frac{{R_0^2}}{32}}+{\frac{17\,{\hat f_{1}}\,{R_0^3}}{288}}
\cp\nonumber\\ && +\at-{\frac{3\,{\hat
f_{0}}}{16}}-{\frac{R_0}{8}}+{\frac{3\,{\hat f_{1}}\,{R_0^2}}{16}}
+{\frac{{\hat f_{2}}\,{R_0^4}}{288}}\ct\,\Delta_0 \nonumber\\ &&\hs
+\at-{\frac{3}{32}} +{\frac{{\hat f_{1}}\,R_0}{32}} +{\frac{{\hat
f_{2}}\,{R_0^3}}{48}}\ct\,\Delta_0^2 \nonumber\\ && \ap\hs\hs\hs
+{\frac{{\hat f_{2}}\,{R_0^2}}{32}}\,\Delta_0^3\cq\,\Delta_0\sigma
\nonumber\\&& +h\,\aq{\frac{R_0}{16}}-{\frac{{\hat
f_{1}}\,{R_0^2}}{16}} -{\frac{{\hat f_{2}}\,{R_0^4}}{72}}
\cp\nonumber\\ &&\hs +\at{\frac{3}{16}}-{\frac{3\,{\hat
f_{1}}\,R_0}{16}} -{\frac{{\hat f_{2}}\,{R_0^3}}{16}}\ct\,\Delta_0
\nonumber\\&&\ap\hs\hs\hs -{\frac{{\hat
f_{2}}\,{R_0^2}}{16}}\,\Delta_0^2\cq\,\Delta_0\sigma\,, \eeq
where
$\lap_0$,  $\lap_1$ and $\lap_2$ are the Laplace-Beltrami operators
acting on scalars, transverse vector and
traceless-transverse tensor fields, respectively. The latter
expression is valid off-shell, that is, for an arbitrary choice of
the function $f(G_{0})$.

As is well-known, invariance under diffeomorphisms renders the
operator related to the latter quadratic form not invertible in the
$(h,\si)$ sector. One needs a gauge fixing term and a corresponding
ghost compensating term.
We can use the same class of gauge conditions chosen in
Ref.~\cite{Cognola:2005de} and for this reason we refere the reader
to that paper, the gauge-fixing ${\cal L}_{gf}$
and ghost ${\cal L}_{gh}$ contributions to the
quadratic Lagrangian ${\cal L}={\cal L}_2+{\cal L}_{gf}+{\cal L}_{gh}$
being the very same, that is
\begin{eqnarray}
{\cal L}_{gf} &=&\frac{\al}2\aq\xi^\mu\,\at\lap_1+\frac{R_0}4\ct^2\,\xi_\mu
\cp\nonumber\\&&\hs
    +\frac{3\rho}{8}\,h\,\at\lap_0+\frac{R_0}3\ct\,\lap_0\,\si
\nonumber\\&&\ap
-\frac{\rho^2}{16}\,h\,\lap_0\,h
-\frac{9}{16}\,\si\,\at\lap_0+\frac{R_0}3\ct^2\,\lap_0\,\si
\cq
\nonumber\\&&
+\frac{\beta}2\aq\xi^k\,\at\lap_1+\frac{R_0}4\ct^2\,\lap_1\xi_k
\cp\nonumber\\&&
     +\frac{3\rho}{8}\,h\,\at\lap_0+\frac{R_{0}}4\ct\at\lap_0
+\frac{R_{0}}3\ct\,\lap_0 \si
\nonumber\\&&\hs
     -\frac{\rho^2}{16}\,h\,\at\lap_0+\frac{R_0}4\ct\,\lap_0 h
\nonumber\\&&\hs
     -\frac{9}{16}\,\si\,\at
    \lap_0+\frac{R_0}4\ct
\nonumber\\&&\hs\hs\times \ap
\at\lap_0+\frac{R_0}3\ct^2\,\lap_0 \si \cq\,,
\label{AAA10} \eeq
\beq {\cal L}_{gh} &=&\alpha\aq\hat B^\mu\at\lap_1+\frac{R_0}{4}\ct\hat C_\mu
\cp\nonumber\\ &&\hs\ap
+\frac{\rho-3}{2}\,b\,\at\lap_0
-\frac{R_0}{\rho-3}\ct\,\lap_0 c\cq
\nonumber\\&&
+\beta\aq\hat B^\mu\,\at\lap_1+\frac{R_0}{4}\ct\,\lap_1\,\hat C_\mu \cp
\nonumber \\ &&\hs
\ap +\frac{\rho-3}{2}\,b\,\at\lap_0+\frac{R_0}{4}\ct
\cp\nonumber\\&&\hs\hs\ap\times
    \at\lap_0-\frac{R_0}{\rho-3}\ct\,\lap_0 c\cq\,,
\eeq
where $\alpha,\beta,\rho$ are arbitrary parameters and
$\hat C^\mu,c$, $\hat B^\mu,b$ are the irreducible components of
ghost and anti-ghost fields.


Now, via standard path integral quantization and zeta-function
regularization \cite{buch,frad,eli94}, one can compute
the one-loop effective action as a determinant of a differential operator,
exactly in the same way as it has been done in Ref.~\cite{Cognola:2005de}.
Here the technical difficulty comes from the fact that the
determinant which gives the one-loop effective action,
also after simplifications, is a polynomial of fifth order in the
Laplace operator and thus, in general, it is not possible to write
it as a product of determinants of Laplace-like operators.
The general structure of the one-loop effective action $\Ga^{(1)}$
is of the form
\beq
e^{-\Ga^{(1)}}&=&\det(\lap_0-R_0)\det\at\lap_1+\frac{R_0}{4}\ct\,
   \nonumber\\ &&\hs\times\,
\det\aq\sum_{k=0}^5\,W_k\De_0^k\cq^{-1/2}
\nonumber\\ &&\times\,
\det\aq(\De_1+Y)(\De_1+Z)\cq^{-1/2}
\nonumber\\ &&\hs\times\,
\det(\De_2+X)^{-1/2}
\label{OLEA}\eeq
where $X,Y,Z,W_k$ are complicated
functions of $\hat f_0,\hat f_1,\hat f_2$. In principle, they can be
exactly computed, but we will not write them explicitly here since,
as they stand, they are not really
convenient for direct applications to the dark energy problem.
However, Eq.~(\ref{OLEA}) is certainly useful for a numerical analysis of
the models, since the eigenvalues of Laplace-like operators on
de Sitter manifolds are exactly known and, as a consequence,
the one-loop effective action can be obtained in closed form.
They could be also useful
for the study of the stability of the de Sitter solutions
of GB modified gravity, as it happens
for the simpler case of ${\cal F}(R)$ models \cite{Cognola:2005sg}.

One can see that the structure of Eq.~(\ref{OLEA}), here written
for the gauge parameter $\beta=0$, is similar to
the one for the analog equation (3.24) in Ref.~\cite{Cognola:2005sg},
the only difference being due to the fact that (\ref{OLEA}) contains
some additional contributions, since the starting classical action
here depends not only on curvature, but also on other invariants.

\section{The transition from the deceleration to the acceleration era in
modified curvature-Gauss-Bonnet gravity}

It is interesting to study late-time cosmology in generalized
theories, which include both the functional dependence from
curvature as well as from the Gauss-Bonnet term (for some
investigation of the related Ricci-squared gravity cosmology, see
\cite{allemandi}). Our starting  action is (\ref{GR1}). In the 
case $\rho_m=0$, we can
consider the situation where $F(G,R)$ has the following form, as an
explicitly solvable example:
\be
\label{GR7}
F(G,R)=R \tilde
f\left(\frac{G}{R^2}\right)\ .
\ee
If we assume that $H=h_0/t$, with
a constant $h_0$, Eq.~(\ref{GR4}) reduces to an algebraic equation:
\bea
\label{GR8}
0&=& 2\left(3h_0 - 1\right)\tilde f'(C) +
3\left(2h_0 -1\right)^2\tilde f(C)\ ,\nn
C&\equiv &
\frac{2h_0\left(h_0 - 1\right)}{3\left(2h_0 - 1\right)^2}\ .
\eea
As a further example, we consider the model where
\be
\label{GR9}
\tilde f\left(\frac{G}{R^2}\right)=\frac{1}{2\kappa^2} + f_0
\left(\frac{G}{R^2}\right)\ .
\ee
Here, it may be shown that
\bea
\label{GR10}
0&=& \left(\frac{6}{\kappa^2} + 2f_0\right)h_0^2 -
2\left(\frac{3}{\kappa^2} - 2f_0\right)h_0 \nn
&& + \frac{3}{2\kappa^2} + 2f_0\ .
\eea
When $f_0<0$ and $f_0>3/8\kappa^2$, Eq.~(\ref{GR10}) has the following
solutions:
\be
\label{GR11}
h_0 = \frac{\frac{3}{\kappa^2} - 2f_0 \pm \sqrt{
8f_0\left(f_0 - \frac{3}{8\kappa^2}\right)}} {\frac{6}{\kappa^2} +
2f_0}\ .
\ee
One can solve (\ref{GR10}) with respect to $\kappa^2f_0$:
\be
\label{GR12}
\kappa^2 f_0 = \tilde G(h_0) \equiv -
\frac{3\left(2h_0 - 1\right)^2}{4\left(h_0^2 + 2h_0 -1\right)}
\ee
and, therefore, by properly choosing $f_0$, we obtain a theory for
any specified $h_0$. It is easy to check that $\tilde G(\infty)=3$,
$\tilde G(1/2)=0$, and $\tilde G\left(-1\pm \sqrt{2}\right)=\infty$. Since
\be
\label{GR13}
{\tilde G}'(h_0) = - \frac{3\left(2h_0 - 1\right)\left(3h_0 -
1\right)}{2\left(h_0^2 + 2h_0 -1\right)^2}\ ,
\ee
we also get that $\tilde G(h_0)$ has extrema for $h_0=1/2$, $1/3$. 
Moreover,
$\tilde G(1/3)=3/8$.
The qualitative behavior of $\kappa^2 f_0=\tilde G(h_0)$ is given in
Fig.~\ref{Fig1}. Then, for $\kappa^2 f_0<-3$, there is a solution
describing a phantom with $h_0<-1-\sqrt{2}$ and a solution
describing effective matter with $h_0>-1+\sqrt{2}$. When
$-3<\kappa^2 f_0<0$, there are two solutions, describing effective
matter with $h_0>-1+\sqrt{2}$. When $0<\kappa^2 f_0<3/8$, there is
no solution. When $3/8<\kappa^2 f_0<3/2$, there are two solutions
describing matter with $0<h_0<-1+\sqrt{2}<1$. When $\kappa^2
f_0>3/2$, there is a solution describing a phantom era with
$-1-\sqrt{2}<h_0<0$ and a solution describing an effective matter
era with $<1/3<h_0<-1+\sqrt{2}$. As one sees, there can be indeed
solutions describing an effective phantom era, in general.

\begin{figure}
\begin{center}
\unitlength=0.6mm
\begin{picture}(120,100)

\thicklines

\put(0,50){\vector(1,0){110}}
\put(60,0){\vector(0,1){90}}
\qbezier(68.1,0)(68.1,50)(70,50)
\qbezier(70,50)(80,50)(85,40)
\qbezier(85,40)(95,20.5)(110,20.5)

\qbezier(67.5,90)(67.5,53.75)(66.67,53.75)
\qbezier(66.67,53.75)(62.1875,53.75)(60,65)
\qbezier(60,65)(55,85)(54,90)

\qbezier(11,0)(11.7,19)(0,19)

\put(115,50){\makebox(0,0){$h_0$}}
\put(60,95){\makebox(0,0){$\kappa^2 f_0$}}

\put(13,51){$-1-\sqrt{2}$}
\put(57,53.75){\makebox(0,0){$\frac{3}{8}$}}
\put(63,65){\makebox(0,0){$\frac{3}{2}$}}
\put(65,46){\makebox(0,0){$\frac{1}{3}$}}
\put(71,46){\makebox(0,0){$\frac{1}{2}$}}
\put(74,56){$-1+\sqrt{2}$}
\put(53,21){$-3$}

\thinlines

\put(0,20){\line(1,0){110}}
\put(11.8,0){\line(0,1){90}}
\put(68,0){\line(0,1){90}}
\put(66.67,50){\line(0,1){3.75}}
\put(60,53.75){\line(1,0){6.14}}

\put(68,50){\line(1,1){5}}

\end{picture}
\end{center}
\caption{The qualitative behavior of $\tilde G(h_0)$.
\label{Fig1}}
\end{figure}
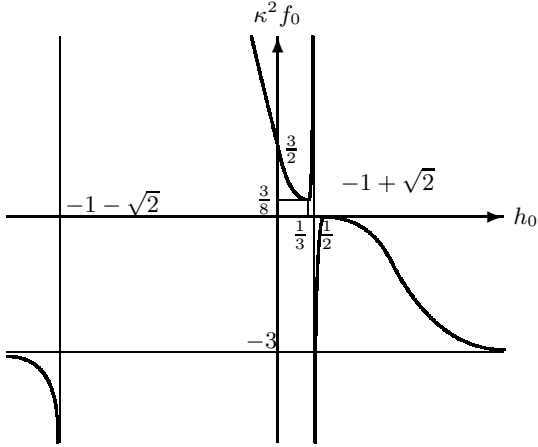

Observational data hint towards the fact that the deceleration of
the universe turned into acceleration about five billion years ago.
We now investigate if we can construct a model describing the
transition from the deceleration phase to the acceleration one, in
the present formulation. To this end, we consider the following
model
\be
\label{GR14}
F(G,R)=R\left(\frac{1}{2\kappa^2} + f_0
\frac{G}{R^2}\right) + \frac{g_0}{G}\ .
\ee
When the curvature is
large, as in the case of the primordial universe, the last term can
be neglected and the model (\ref{GR14}) reduces to (\ref{GR9}).
Then, with the choice $3/8<\kappa^2 f_0<3/2$, it follows
$0<h_0<-1+\sqrt{2}<1$ and therefore the universe is decelerating. If
the curvature becomes small, as in the present universe, the last
term becomes large. But, by including the last term, we have a
solution describing a de Sitter universe in (\ref{GR6}), which has
the form:
\be
\label{GR15}
0=-\frac{H_0^2}{\kappa^2}\left(\kappa^2
f_0 + 3\right) - \frac{g_0}{12H_0^4}\ .
\ee
As $\kappa^2 f_0 + 3$ is
positive when $3/8<\kappa^2 f_0<3/2$, Eq.~(\ref{GR16}) has a real
solution for $g_0<0$:
\be
\label{GR16}
H_0=\left\{ - \frac{\kappa^2
g_0}{12\left(\kappa^2 f_0 + 3\right)}\right\}^{1/6}\ .
\ee
Then, the decelerating universe can indeed turn to a de Sitter universe, which
is accelerating. Therefore, the model (\ref{GR14}) could perfectly
describe the transition from deceleration to acceleration. More
complicated, dark energy cosmologies may be constructed in frames of
such theory.

\section{Hierarchy problem in modified gravity}

Recently the hierarchy problem has been investigated, in \cite{BN},
by using a scalar-tensor theory. Here, we will give a somewhat similar but
seemingly more natural models in the scalar-tensor family with
modified gravity. In \cite{BN}, in order to generate the hierarchy,
a small scale, which is the vacuum decay rate $\Gamma_{vac}$, was
considered. Instead of $\Gamma_{vac}$, we here use the age of the
universe, $\sim 10^{-33}\,{\rm eV}$, as the small mass scale.

The following scalar-tensor theory \cite{ENO} can be considered,
as an example
   \bea
\label{P1} S&=&{1 \over \kappa^2}\int d^4x
\sqrt{-g}\e^{\alpha\phi}\Bigl(R - {1 \over 2}
\partial_\mu \phi \partial^\mu \phi \nn
&& - V_0\e^{-2\phi/\phi_0})\Bigr) \nn
&& + \int d^4 x \sqrt{-g}\left( -{1 \over 2}\partial_\mu \chi
\partial^\mu \chi - U(\chi)\right) \ ,
\eea
where $\alpha$, $V_0$, and $\phi_0$ are constant parameters and
$U(\chi)$ is the potential for $\chi$. As the matter scalar, $\chi$,
does not couple with $\phi$ directly, the equivalence principle is
not violated, although the effective gravitational coupling depends
on $\phi$ as
\be
\label{PP0} \tilde\kappa=\kappa\e^{-{\alpha \phi
\over 2}}\ .
\ee
When $\chi=0$,  the  following FRW solution exists
\bea
\label{PP1} && a(t)= a_0
\left(\frac{t}{t_0}\right)^{h_0}\ ,\quad \phi = \phi_0\ln
\frac{t}{t_0}\ ,\nn && h_0\equiv \frac{2\alpha^2\phi_0^2 + \phi_0^2
- 2\alpha\phi_0}{2\left(\alpha\phi_0 + 2\right)}\ ,\nn
&& t_0\equiv \frac{\phi_0^2}{\left(\alpha\phi_0 + 2\right)^2} \\
&& \quad \times \sqrt{\frac{\left(\alpha^2 +
\frac{1}{3}\right)\left(24\alpha^2\phi_0^2 - 4\alpha \phi_0 +
9\phi_0^2 -4\right)}{2V_0}}\ .\nonumber
\eea
The effective EoS parameter
is
\be
\label{PP3} w_{\rm eff}= - 1 + \frac{4\left(\alpha\phi_0 +
2\right)}{3\left(2\alpha^2\phi_0^2 + \phi_0^2 -
2\alpha\phi_0\right)}\ ,
\ee
which can be less than $-1$, in general.

We may perfectly assume the dimensionful parameters $\kappa$ and
$V_0$, and therefore $t_0$ in (\ref{P1}), could be the scale of the
weak interaction $\sim 10^2\,{\rm GeV}=10^{11}\,{\rm eV}$. If
    $t$ is of the order of the age of the universe, $\sim
10^{-33}\,{\rm eV}$, Eq.~(\ref{PP1}) gives
\be
\label{PP4}
\e^{-\alpha\phi/2}\sim 10^{-22\alpha\phi_0}\ .
\ee
Then, if
\be
\label{PP5}
\alpha\phi_0 = \frac{17}{22}\ ,
\ee
$\tilde \kappa$
(\ref{PP0}) is of the order of the Planck length
$\left(10^{19}\,{\rm GeV}\right)^{-1}$. Therefore, using a model
whose action is given by (\ref{P1}), the important hierarchy problem
    might be solved. By substituting (\ref{PP5}) into (\ref{PP3}), one
obtains
\be
\label{PP6}
w_{\rm eff}= - 1 + \frac{61}{33\left(\phi_0^2
- \frac{85}{242}\right)}\ .
\ee
It is seen that $w_{\rm eff}$ can be
less than $-1$ if $\phi_0^2 < {85}/{242}$, and $t_0$ is given by
\be
\label{PP7}
t_0= \frac{22^2
\phi_0^2}{61^2}\sqrt{\frac{\left(\frac{17^2}{22^2\phi_0^2} +
\frac{1}{3}\right)\left(\frac{876}{121} + 9\phi_0^2\right)}{2V_0}}\ ,
\ee
which is  real and positive, as far as $V_0>0$.
Similar method can be applied in the solution of the hierarchy problem in
a generalized scalar-tensor theory including a non-minimal coupling with 
the curvature \cite{faraoni}.

We now start from the following action for modified gravity coupled
with matter
\be
\label{RR1}
S=\int d^4 x \sqrt{-g} \left({1 \over \kappa^2}{\cal F}(R) + {\cal L}_{\rm
matter}\right)\ ,
\ee
${\cal F}(R)$ being some arbitrary function.
   Introducing the auxiliary fields, $A$ and
$B$, one can rewrite the action (\ref{RR1}) as follows:
\be
\label{RR2}
S=\int d^4 x \sqrt{-g} \left[{1 \over
\kappa^2}\left\{B\left(R-A\right) + {\cal F}(A)\right\} + {\cal L}_{\rm
matter}\right]\ .
\ee
One is able to eliminate $B$,  and obtain
\bea
\label{RR6}
S&=&\int d^4 x \sqrt{-g} \Bigl[{1 \over
\kappa^2}\left\{{\cal F}'(A)\left(R-A\right) + {\cal F}(A)\right\} \nn
&& + {\cal L}_{\rm matter}\Bigr]\ ,
\eea
and by using the conformal transformation
\be
\label{RR7}
g_{\mu\nu}\to \e^\sigma g_{\mu\nu}\ ,
\ee
with
\be
\label{RR8b}
\sigma = -\ln {\cal F}'(A)\ ,
\ee
the action (\ref{RR6}) is rewritten as the Einstein-frame action:
\bea
\label{RR10}
S_E&=&\int d^4 x \sqrt{-g} \Bigl[{1 \over
\kappa^2}\left( R - {3 \over 2}g^{\rho\sigma}
\partial_\rho \sigma \partial_\sigma \sigma - V(\sigma)\right) \nn
&& + {\cal L}_{\rm matter}^{\sigma}\Bigr]\ .
\eea
Here,
\bea
\label{RR11}
V(\sigma) &=& \e^\sigma {\cal G}\left(\e^{-\sigma}\right) -
\e^{2\sigma} {\cal F}\left({\cal G}\left(\e^{-\sigma} \right)\right) \nn
&=& {A \over {\cal F}'(A)} - {{\cal F}(A) \over {\cal F}'(A)^2}\ ,
\eea
and we solve (\ref{RR8b}) with respect to $A$ as $A={\cal G}(\sigma)$.
Let us assume that the
matter Lagrangian density ${\cal L}_{\rm matter}$ contains a
Higgs-like scalar field $\varphi$
\be
\label{RR12}
{\cal L}_{\rm matter}= - \frac{1}{2}\nabla^\mu \varphi \nabla_\mu \varphi +
\frac{\mu^2}{2}\varphi^2 - \lambda \varphi^4 + \cdots \ .
\ee
Under the conformal transformation (\ref{RR7}), the matter
Lagrangian density ${\cal L}_{\rm matter}$ is transformed as
\bea
\label{RR12a} {\cal L}_{\rm matter} \to {\cal L}_{\rm matter}^\sigma
&=&  - \frac{\e^{\sigma}}{2}\nabla^\mu \varphi \nabla_\mu \varphi +
\frac{\mu^2\e^{2\sigma}}{2}\varphi^2 \nn
&& - \lambda \e^{2\sigma} \varphi^4 +\cdots \ .
\eea
By redefining $\varphi$ as
\be
\label{RR12b}
\varphi\to \e^{-\sigma/2}\varphi\ ,
\ee
${\cal L}_{\rm matter}^\sigma$ acquires the following form
\be
\label{RR12c} {\cal L}_{\rm matter}^\sigma \sim  - \frac{1}{2}\nabla^\mu
\varphi
\nabla_\mu \varphi + \frac{\mu^2\e^{\sigma}}{2}\varphi^2
    - \lambda \varphi^4 +\cdots \ ,
\ee where  the time derivative of $\sigma$ is neglected. Then, the
massive parameter $\mu$, which  determines the weak scale, is
effectively transformed as \be \label{RR13} \mu\to \tilde \mu\equiv
\e^{\sigma/2}\mu\ . \ee In principle, $\mu$ can be of the order of
the Planck scale $10^{19}$ GeV, but if $\e^{\sigma/2}\sim 10^{-17}$
in the present universe, $\tilde \mu$ could be $10^2$ GeV, which is
the scale of the weak interaction. Therefore, there is a quite
natural possibility that the hierarchy problem can be solved by using the
above version of modified gravity.

We may consider the model \be \label{RR14} {\cal F}(R) = R + f_0 R^\alpha\
, \ee with constant $f_0$ and $\alpha$. If $\alpha<1$, the second
term dominates, when the curvature is small. Assuming that the EoS
parameter, $w$, of matter is constant one gets \cite{ANO} \bea
\label{M8} && a=a_0 t^{h_0} \ ,\quad h_0\equiv
\frac{2\alpha}{3(1+w)} \ ,\nn && a_0\equiv
\Bigl[-\frac{6f_0h_0}{\rho_0}\left(-6h_0 + 12
h_0^2\right)^{\alpha-1} \nn &&
\times\left\{\left(1-2\alpha\right)\left(1-\alpha\right) -
(2-\alpha)h_0\right\} \Bigr]^{-\frac{1}{3(1+w)}}\ , \eea and, by
using (\ref{M8}), we find the effective $w_{\rm eff}$ to be given by
\be \label{M9} w_{\rm eff}=-1 + \frac{1+w}{\alpha}\ . \ee Hence, if
$w$ is larger than $-1$ (as for effective quintessence or even for a
usual ideal fluid with positive $w$), when $\alpha$ is negative,
    an effective phantom phase occurs where $w_{\rm eff}$ is less than
$-1$. Note that this is different from the case of pure modified
gravity.

By using (\ref{RR8b}) and neglecting the first term in (\ref{RR14}),
it follows that \be \label{M10} \e^{\sigma/2}\sim \frac{1}{\sqrt{f_0
\alpha R^{\alpha-1}}}\ . \ee In the present universe,  $R\sim
\left(10^{-33}\,{\rm eV}\right)^2$. Assume now that  $f_0$ could be
given by the Planck scale $\sim 10^{19}\,{\rm GeV}=10^{28}\,{\rm
eV}$ as $f_0\sim \left(10^{28}\,{\rm
eV}\right)^{\frac{1}{2(\alpha-1)}}$. Then, Eq.~(\ref{M10}) would
yield \be \label{M11} \e^{\sigma/2}\sim 10^{61\left(\alpha
-1\right)}\ . \ee If we furthermore assume that $-17=61\left(\alpha
-1\right)$, we find $\alpha = 44/61$. In that case, if $w>-1$,
$w_{\rm eff}>-1$ and the universe is not phantom like, but
(\ref{M11}) hints towards the possibility that modified gravity can
solve in fact the hierarchy problem.

Let us write the action of the scalar-tensor theory as
\bea
\label{F1}
S&=&{1 \over \kappa^2}\int d^4x
\sqrt{-g}\e^{\alpha\phi}\Bigl(R - {1 \over 2}
\partial_\mu \phi \partial^\mu \phi \nn
&& - V_0\e^{-2\phi/\phi_0})\Bigr) \nn
&& + \int d^4 x \sqrt{-g} {\cal L}_{\rm matter} \ , \nn
{\cal L}_{\rm matter} &=& - \frac{1}{2}\partial^\mu \varphi \partial_\mu
\varphi + \frac{\mu^2}{2}\varphi^2
    - \lambda \varphi^4 + \cdots \ .
\eea
This action could be regarded as the Jordan frame action.
Scalar field $\varphi$ could be identified with the Higgs field in the
weak(-electromagnetic)
interaction.
Now the  ratio of the inverse of the the effective gravitational coupling
$\tilde\kappa=\kappa\e^{-{\alpha \phi \over 2}}$  (\ref{PP0})
and the Higgs mass  $\mu$ is given by
\be
\label{F2}
\frac{\frac{1}{\tilde\kappa}}{\mu}=\frac{\e^{{\alpha \phi \over
2}}}{\kappa\mu}\ .
\ee
Hence, even if both of $1/\kappa$ and $\mu$ are of the order of the
    weak interaction scale,
if $\e^{{\alpha \phi \over 2}}\sim 10^{17}$, $1/\tilde\kappa$ could
be of the order of the Planck scale.

By rescaling the metric and the Higgs scalar  $\varphi$ as
\be
\label{F3}
g_{\mu\nu}\to \e^{-\alpha \phi} g_{\mu\nu}\ ,\quad
\varphi \to \e^{{\alpha \phi \over 2}}\varphi\ ,
\ee
    the Einstein frame action is obtained:
\bea
\label{F4}
S&=&{1 \over \kappa^2}\int d^4x
\sqrt{-g}\Bigl(R - {1 \over 2}\left(1 + 3\alpha^2\right)
\partial_\mu \phi \partial^\mu \phi \nn
&& - V_0\e^{-2\phi\left(1/\phi_0+\alpha\right)})\Bigr) \nn
&& + \int d^4 x \sqrt{-g} {\cal L}_{\rm matter} \ , \nn
{\cal L}_{\rm matter} &=& - \frac{1}{2}
\left(\partial^\mu \varphi + \frac{\alpha}{2}\partial^\mu\phi \varphi\right)
\left(\partial_\mu \varphi + \frac{\alpha}{2}\partial_\mu\phi \varphi\right)
\nn
&& + \frac{\mu^2\e^{-\alpha \phi}}{2}\varphi^2
    - \lambda \varphi^4 + \cdots \ .
\eea
In the Einstein frame, the gravitational coupling $\kappa$ is constant
    but the effective Higgs mass,
$\tilde \mu$, defined by \be \label{F5} \tilde\mu \equiv
\e^{-{\alpha \phi \over 2}}\mu \ , \ee can be time-dependent. Hence,
the ratio of $1/\kappa$ and $\tilde\mu$ is given by \be \label{F6}
\frac{\frac{1}{\kappa}}{\tilde\mu}=\frac{\e^{{\alpha \phi \over
2}}}{\kappa\mu}\ , \ee which is identical with (\ref{F2}). Then even
if both of $1/\kappa$ and $\mu$ are of the order of the Planck
scale, if $\e^{{\alpha \phi \over 2}}\sim 10^{17}$, $\tilde\mu$
could be an order of  the weak interaction scale. Therefore the
solution of the hierarchy problem does not essentially depend on the
   choice of frame.

Nevertheless, note that the cosmological time variables in the two
frames could be different due to the scale transformation
(\ref{F3}), as \be \label{F7} dt \to d\tilde t = \e^{-{\alpha \phi
\over 2}}dt\ . \ee Therefore, the time-intervals are different in
the two frames. The units of time and length are now defined by
electromagnetism. Then, the frame where the electromagnetic fields
do not couple with the scalar field $\phi$ could be physically more
preferable. Since the electromagnetic interaction is a part of the
electro-weak interaction, the Jordan frame in (\ref{F1}) should be
more preferable from the point of view of the solution of the
hierarchy problem.

In the case of $f(G)$-gravity, whose action is given by (\ref{GR1}) with
(\ref{ff1}),
it is rather difficult to solve the hierarchy problem in the same
way, since the factor in front of the scalar curvature, which should
be the inverse of the Newton constant, although it is indeed a
constant, in the above cases this factor depends on time. However,
when including a term like $g(G)R$, where $g(G)$ is a proper
function of the Gauss-Bonnet invariant, the effective Newton
constant could become time-dependent and might indeed help solve the
hierarchy problem.


Similar mechanism can work also in $F(G,R)$-gravity  (\ref{GR1}).
Introducing the auxiliary fields, $A$, $B$, $C$, and $D$,
one can rewrite the action (\ref{RR1}) as follows:
\bea
\label{GRR2}
S&=&\int d^4 x \sqrt{-g} \Bigl[\frac{1}{\kappa^2}\{B\left(R-A\right) + D\left(G
- C\right) \nn
&& + F(A,C)\} + {\cal L}_{\rm matter}\Bigr]\ .
\eea
One is able to eliminate $B$ and $D$,  and obtain
\bea
\label{GRR6}
S&=&\int d^4 x \sqrt{-g} \Bigl[\frac{1}{\kappa^2}\{\partial_A
F(A,C)\left(R-A\right) \nn
&& + \partial_C F(A,C)\left(G-C\right) + F(A,C)\} \nn
&& + {\cal L}_{\rm matter}\Bigr]\ .
\eea
If  a scalar field is deleted  by
\be
\label{GRRa1}
\e^{-\sigma}\equiv \partial_A F(A,C)\ ,
\ee
one can solve (\ref{GRRa1}) with respect to $A$ as $A=A(\sigma, C)$.
Then, we obtain
\bea
\label{GRRa2}
S&=&\int d^4 x \sqrt{-g}
\Bigl[\frac{1}{\kappa^2}\{\e^{-\sigma}\left(R-A(\sigma,C)\right) \nn
&& + \partial_C F(A(\sigma,C),C)\left(G-C\right) + F(A(\sigma,C),C)\} \nn
&& + {\cal L}_{\rm matter}\Bigr]\ .
\eea
Varying over $C$, it follows that $C=G$, which allows to eliminate $C$
\bea
\label{GRRa3}
S&=&\int d^4 x \sqrt{-g}
\Bigl[\frac{1}{\kappa^2}\{\e^{-\sigma}\left(R-A(\sigma,G)\right) \nn
&& + F(A(\sigma,G),G) \} + {\cal L}_{\rm matter}\Bigr]\ .
\eea
Performing the scale transformation (\ref{RR7}), we obtain the 
Einstein frame action.
If we consider matter (Higgs) scalar as in (\ref{RR12}), we can redefine
$\varphi$ as in (\ref{RR12b}).
The same scenario as in the case of ${\cal F}(R)$-gravity is applied, if
$\e^{\sigma/2}\sim 10^{-17}$
in the present universe, there is a possibility that the hierarchy
 problem can be solved by working in the above version of $F(G,R)$ gravity.
Hence, we have proven that the hierarchy problem can indeed be solved in
modified gravity which contains a $F(G, R)$ term.

\section{Discussion}

To summarize, various types of dark energy cosmologies in modified
Gauss-Bonnet gravity ---which can be viewed as inspired by string
considerations--- have been investigated in this paper. We have shown, 
in particular, that effective
quintessence, phantom and cosmological constant eras can naturally emerge
in this framework, without the need to introduce scalar fields of
any sort explicitly. Actually, the cosmic acceleration we observe
 may result from the
expansion of the universe due to the growing of the extra terms in
the gravitational action when the curvature decreases. In addition,
with the help of several examples, corresponding to explicit choices of the
function $f(G)$, we have shown that the unification of early-time
inflation with late-time acceleration in those theories occurs also quite
naturally. Moreover, the framework is  attractive in the sense
that it leads to a  reasonable behavior in the Solar System
limit (no corrections to Newton's law, no instabilities, no
Brans-Dicke problems appear), whatever be the particular choice of
$f(G)$. Finally, the transition from the deceleration to the
acceleration epoch, or from the non-phantom to the phantom regime
---provided the current universe is in its phantom phase!--- may both be
natural ingredients of our theory, without the necessity to invoke any
sort of exotic matter (quintessence or phantom) with an explicit
negative EoS parameter.

We have also shown in the paper that modified GB gravity has de
Sitter or SdS BH solutions, for which the corresponding entropy has
been calculated. It has been explicitly demonstrated that our theory
can be consistently  quantized to one-loop order in de Sitter space,
in the same way as modified ${\cal F}(R)$ gravity.

Dark energy cosmologies in a more complicated $F(G,R)$ framework can
be constructed in a similar fashion too. An attempt to address
fundamental particle physics issues (as  the hierarchy
problem), as resulting from a modification of gravity, has shown
that some natural solution may possibly be achieved in
${\cal F}(R)$-gravity, but probably not in $f(G)$-gravity (albeit the case
$F(G,R)$ opens again a new possibility)). In this respect, it may also be
of interest to study other modified gravities, where a non-minimal
coupling of the sort $L_d$-$f(G)$, with $L_d$ being some matter
Lagrangian which includes also a kinetic term is introduced. In the
specific case of a $L_d$-${\cal F}(R)$ non-minimal coupling, such terms may
be able to explain the current dark energy dominance\cite{dominance}
as a gravitationally-assisted one.

The next step should be to fit the specific astrophysical
predictions of the above theory with current observational data (for
a recent summary and comparison of such data from various sources,
see \cite{padmanabhan}), which ought to be modified accordingly, as
most of them are derived under the (often implicit) assumption that
standard General Relativity is correct. One immediate possibility is
to study the perturbation structure in close analogy with what has
been done for ${\cal F}(R)$-gravity \cite{inh,pert}. This will be reported
elsewhere.

\section*{Acknowledgments}
We thank L. Vanzo for useful discussions. Support from the AI
program INFN(Italy)-CICYT(Spain), from MEC (Spain), project
FIS2005-01181 (SDO) and SEEU grant PR2004-0126 (EE), is gratefully
acknowledged.

\appendix

\section{}

Having in mind the importance of spherically symmetric BH solutions in
gravity theories, let us consider the possibility that $F(G,R)$-gravity 
has the Schwarzschild black hole solution:
\bea
\label{GBs2}
&& ds^2 = - \e^{2\nu}dt^2 + \e^{-2\nu}dr^2 + r^2 d\Omega_2^2\ ,\nn
&& \e^{2\nu}= 1 - \frac{r_0}{r}\ ,
\eea
For simplicity, we consider the vacuum case ($T_{\mu\nu}=0$) and
  concentrate on the model (\ref{ff1}).
  Then, by multiplying $g_{\mu\nu}$ with (\ref{GR2}) for the action
(\ref{ff1}),
we find
\bea
\label{GBs1}
0 &=& \frac{1}{2\kappa^2}R + 2 f(G)
-2 f'(G) R^2 + 4f'(G)R_{\mu\rho} R^{\mu\rho} \nn
&& -2 f'(G) R^{\mu\rho\sigma\tau}R_{\mu\rho\sigma\tau}
+4 f'(G) R^{\mu\nu}R_{\mu\nu} \nn
&& -2 \left( \nabla^2 f'(G)\right)R
+4 \left( \nabla_\nu \nabla^\mu f'(G)\right)R^{\nu\mu} \ .
\eea
In case of the Schwarzschild black hole (\ref{GBs2}), one has
\bea
\label{GBs3}
&& R=R_{\mu\nu}=0\ ,\nn
&& G=R_{\mu\nu\rho\sigma} R^{\mu\nu\rho\sigma}=\frac{12r_0^2}{r^6}\ .
\eea
Then, Eq.(\ref{GBs1}) is reduced to be
\be
\label{GBs4}
0= f(G) - G f'(G)\ ,
\ee
which gives
\be
\label{GBs5}
f(G)=f_0 G\ ,
\ee
with a constant $f_0$.
Since $G$ is a total derivative, one can drop $f(G)$ in (\ref{GR1}) with
(\ref{ff1}) for
(\ref{GBs5}).
Hence, the Schwarzschild black hole geometry is not a solution for a
non-trivial $f(G)$-gravity which justifies our interest for SdS BH in
section III as for the spherically symmetric solution
  of the above $f(G)$ theory.
Note that a theory of this sort may contain a Schwarzschild solution
  in higher dimensions, where $G$ is not a topological invariant
  (for a recent example, see \cite{TM}).
It should be also stressed that modified gravity of the
$F(G,R)$ form containing more complicated $R$-dependent terms
might admit the standard Schwarzschild black hole solution.


\begin{thebibliography}{99}

\bibitem{sami}
E.~Copeland, M.~Sami and S.~Tsujikawa,
``Dynamics of dark energy'', to appear.

\bibitem{pad}
T.~Padmanabhan, Phys.\ Repts.\ {\bf 380}, 235 (2003).

\bibitem{capozziello}
S.~Capozziello,
Int.\ J.\ Mod.\ Phys. D {\bf 11}, 483 (2002);

S.~Capozziello, S.~Carloni and A.~Troisi,
arXiv:astro-ph/0303041;

S.~M.~Carroll, V.~Duvvuri, M.~Trodden and M.~S.~Turner,
Phys.\ Rev.\ D {\bf 70}, 043528 (2004)
[arXiv:astro-ph/0306438].

\bibitem{NOPRD}
S.~Nojiri and S.~D.~Odintsov,
Phys.\ Rev.\ D {\bf 68}, 123512 (2003),
[arXiv:hep-th/0307288].

\bibitem{ln}
S.~Nojiri and S.~D.~Odintsov
GRG {\bf 36}, 1765 (2004),
[arXiv:hep-th/0308176];

X.~Meng and P.~Wang,
Phys.\ Lett.\ B {\bf 584}, 1 (2004),
[arXiv:hep-th/0309062].

\bibitem{tr}
D.~Easson, F.~Schuller, M.~Trodden and M.~Wohlfarth,
arXiv:astro-ph/0506392.

\bibitem{string}
S.~Nojiri, S.~D.~Odintsov and M.~Sasaki,
Phys.\ Rev.\ D {\bf 71}, 123509 (2005)
[arXiv:hep-th/0504052];

M.~Sami, A.~Toporensky, P.~Trejakov and S.~Tsujikawa,
Phys.\ Lett.\ B {\bf 619}, 193 (2005)
[arXiv:hep-th/0504154];

G.~Calcagni, S.~Tsujikawa and M.~Sami,
Class.\ Quant.\ Grav.\ {\bf 22}, 3977 (2005)
[arXiv:hep-th/0505193];

B.~Carter and I.~Neupane
arXiv:hep-th/0510109; 
arXiv:hep-th/0512262.

\bibitem{cai}
H.~Wei and R.~Cai
arXiv:astro-ph/0512018.

\bibitem{CNO}
S.~Capozziello, S.~Nojiri and S.~D.~Odintsov
arXiv:hep-th/ 0512118.

\bibitem{multamaki}
T.~Multamaki and I.~Vilja
arXiv:astro-ph/0506692.

\bibitem{GB}
S.~Nojiri and S.~D.~Odintsov,
Phys.\ Lett.\ B {\bf 631}, 1 (2005)
[arXiv:hep-th/0508049];

S.~Nojiri, S.~D.~Odintsov and O.~G.~Gorbunova,
arXiv:hep-th/0510183.

\bibitem{TM}
T.~Torii, H.~Maeda,
Phys.\ Rev.\ D {\bf 72}, 064007 (2005)
[arXiv:hep-th/0504141];

\bibitem{BT}
T.~Clifton, J.~D.~Barrow,
arXiv:gr-qc/0601118;
arXiv:gr-qc/0511076.

\bibitem{brett}
B.~McInnes, JHEP {\bf 0208}, 029 (2002)
[arXiv:hep-th/0112066].

\bibitem{tsujikawa}
S.~Nojiri, S.~D.~Odintsov and S.~Tsujikawa,
Phys.\ Rev.\ {\bf D71}, 063004 (2005).

\bibitem{ANO}
M.~C.~B.~Abdalla, S.~Nojiri and S.~D.~Odintsov,
Class.\ Quant.\ Grav. {\bf 22}, L35 (2005)
[arXiv:hep-th/0409177].

\bibitem{DK}
A.~D.~Dolgov and M.~Kawasaki, 
Phys.\ Lett.\ B {\bf 573}, 1 (2003)
[arXiv:astro-ph/0307285];

M.~E.~Soussa and R.~P.~Woodard, 
GRG {\bf 36}, 855 (2004)
[arXiv:astro-ph/0308114];

S.~Nojiri and S.~D.~Odintsov, 
Mod.\ Phys.\ Lett.\ A {\bf 19}, 627 (2004)
[arXiv:hep-th/0310045].

\bibitem{newton}
S.~Capozziello,
arXiv:gr-qc/0412088;

X.~Meng and P.~Wang, 
GRG {\bf 36}, 1947 (2004);

A.~Dominguez and D.~Barraco, 
Phys.\ Rev. {\bf D70}, 043505 (2004);

G.~Allemandi, M.~Francaviglia, M.~Ruggiero and A.~Tartaglia,
arXiv:gr-qc/0506123;

T.~Koivisto,
arXiv:gr-qc/0505128;

G.~Olmo,
Phys.\ Rev. {\bf D72}, 083505 (2005);

I.~Navarro and K.~Acoleyen,
arXiv:gr-qc/0506096;

J.~Cembranos,
arXiv:gr-qc/0507039;

T.~Sotirou,
arXiv:gr-qc/0507027;

N.~Poplawski,
arXiv:gr-qc/0510007;

M.~Amarzguioui, O.~Elgaroy, D.~F.~Mota and T.~Multamaki,
arXiv:astro-ph/0510519;

C.~Shao, R.~Cai, B.~Wang and R.~Su,
arXiv:gr-qc/0511034;

S.~Capozziello and A.~Troisi,
arXiv:astro-ph/0507545.

\bibitem{inh}
S.~Nojiri and S.~D.~Odintsov, 
Phys.\ Rev. D {\bf 72}, 023003 (2005),
[arXiv:hep-th/0505215];

S.~Capozziello, V.~F.~Cardone, E.~Elizalde, S.~Nojiri and S.~D.~Odintsov,
arXiv:astro-ph/0508350.

\bibitem{barrow}
J.~Barrow, 
Nucl.\ Phys.\ B {\bf 310}, 743 (1988);

I.~Brevik and O.~Gorbunova,
arXiv:gr-qc/0504001;

M.~Giovannini, 
Phys.\ Lett.\ B {\bf 622}, 349 (2005);

M.~Hu and X.~Meng,
arXiv:astro-ph/0511615;
arXiv:astro-ph/0511163.

\bibitem{st}
S.~Nojiri and S.~D.~Odintsov, 
Phys.\ Rev.\ D {\bf 70}, 103522 (2004),
[arXiv:hep-th/0408170];

H.~Stefancic, 
Phys.\ Rev.\ D {\bf 71}, 084024 (2005);

J.~S.~Alcaniz and H.~Stefancic,
arXiv:astro-ph/0512622.

\bibitem{ginspar}
P.~Ginsparg and M.~J.~Perry,
Nucl.\ Phys.\ B {\bf 222}, 245 (1983).

\bibitem{wipf}
M.~S.~Volkov and A.~Wipf. 
Nucl.\ Phys.\ B {\bf 582}, 313 (2000).

\bibitem{brevik}
I.~Brevik, S.~Nojiri, S.~D.~Odintsov and L.~Vanzo,
Phys.\ Rev.\ D {\bf 70}, 043520 (2004)
[arXiv:hep-th/0401073].

\bibitem{wald}
R.~M.~Wald
Phys.\ Rev.\ D {\bf 48}, (1993) 3427,
[arXiv:gr-qc/9307038].

\bibitem{frad}
E.~S.~Fradkin and A.~A.~Tseytlin,
Nucl.\ Phys.\ B {\bf 234}, (1984) 472.

\bibitem{buch}
I.~L.~Buchbinder, S.~D.~Odintsov and I.~L.~Shapiro,
{\em Effective action in quantum gravity},
IOP Publishing, Bristol, (1992).

\bibitem{eli94}
E.~Elizalde, S.~D.~Odintsov, A.~Romeo, A.~A.~Bytsenko and S.~Zerbini
{\em Zeta regularization techniques with applications}, 
World Scientific, (1994);

E.~Elizalde, 
{\em Ten physical applications of spectral zeta functions}, 
Lecture Notes in Physics, Springer-Verlag, Berlin,
(1995);

A.~A.~Bytsenko, G.~Cognola, E.~Elizalde, V.~Moretti and S.~Zerbini,
{\em Analytic aspects of quantum fields}, 
World Scientific, Singapore, (2003).

\bibitem{Cognola:2005de}
G.~Cognola, E.~Elizalde, S.~Nojiri, S.~D.~Odintsov and S.~Zerbini,
JCAP {\bf 0502}, 010 (2005)
[arXiv:hep-th/0501096].

\bibitem{Cognola:2005sg}
G.~Cognola and S.~Zerbini,
arXiv:hep-th/0511233.

\bibitem{allemandi}
G.~Allemandi, A.~Borowiec and M.~Francaviglia,
Phys.\ Rev.\ D {\bf 70}, 103503 (2004)
[arXiv:hep-th/0407090].

\bibitem{BN}
T.~Biswas and  A.~Notari,
arXiv:hep-ph/0511207.

\bibitem{ENO}
E.~Elizalde, S.~Nojiri and S.~D.~Odintsov,
Phys.\ Rev.\ D {\bf 70}, 043539 (2004)
[arXiv:hep-th/0405034 ].

\bibitem{faraoni}
V.~Faraoni, 
Ann.\ Phys.\ {\bf 317}, 366 (2005)
[arXiv:gr-qc/0502015];

V.~Faraoni and S.~Nadeau
arXiv:gr-qc/0511094.

\bibitem{padmanabhan}
H.~Jassal, J.~Bagla and T.~Padmanabhan,
arXiv:astro-ph/0506748.

\bibitem{dominance}
S.~Nojiri and S.~D.~Odintsov,
Phys.\ Lett.\ B {\bf 599}, 137 (2004)
[arXiv:astro-ph/0403622];

G.~Allemandi, A.~Borowiec, M.~Francaviglia and S.~D.~Odintsov,
Phys.\ Rev.\ D {\bf 72}, 063505 (2005)
[arXiv:gr-qc/0504057].

\bibitem{pert}
T.~Koivisto and H.~Kurki-Suonio,
arXiv:astro-ph/0509422.

\end{thebibliography}
\end{document}